\documentclass[
reprint,
bibnotes,
amsmath,amssymb,
aps,
prd,
unsortedaddress,floatfix
]{revtex4-2}

\usepackage{aas_macros}
\usepackage{titletoc}

\usepackage{amssymb,amsmath,mathtools}
\usepackage{amsthm}
\usepackage{titletoc}
\usepackage{graphicx}
\usepackage{afterpage}
\usepackage{dcolumn}
\usepackage{enumitem}
\usepackage{bm}
\usepackage{float}
\usepackage[many]{tcolorbox}
\usepackage{shuffle}
\usepackage{xcolor}
\usepackage{hyperref}
\usepackage{physics}
\usepackage{tabularx}
\usepackage{diagbox}
\usepackage{subcaption}
\usepackage{multirow}
\usepackage{caption}

\usepackage{tikz}
\usepackage{tikz-feynman}
\usetikzlibrary{decorations.pathmorphing}
\usetikzlibrary{arrows}
\usepackage{circuitikz}
\usetikzlibrary{arrows.meta}
\usetikzlibrary{shapes.geometric}
\usetikzlibrary{shapes.misc}
\usetikzlibrary{positioning}
\usetikzlibrary{decorations.markings}
\usetikzlibrary{intersections}
\usetikzlibrary{positioning}
\usetikzlibrary{calc,angles,quotes}
\usepackage{verbatim}
\usepackage{caption}
\tikzset{
    vector/.style={
        decoration={snake, aspect=0.75, mirror, segment length=2mm},
        decorate
    },
    photon/.style={decorate, decoration={snake, amplitude=1pt, segment length=4pt}}
}
\tikzset{
  cross/.style={
    path picture={
      \draw
        (path picture bounding box.south west) --
        (path picture bounding box.north east)
        (path picture bounding box.north west) --
        (path picture bounding box.south east);
    }
  }
}

\newlength{\Rwidth}
\newlength{\Rheight}

\usepackage{CJK}
\definecolor{Maroon}{RGB}{128, 0, 0}
\definecolor{Green}{rgb}{0.0, 0.5, 0.0}
\definecolor{Purple}{rgb}{0.5, 0.0, 0.5}
\definecolor{Blue}{rgb}{0.0, 0.0, 0.6}

\begin{document}

\title{The Very Nearly  
\texorpdfstring{\raisebox{-0.52em}{%
  \resizebox{3cm}{!}{\begin{tikzpicture}[scale=0.9,
    line cap=round,
    line join=round,
    vleft/.style={
        circle,
        fill=blue!65!black,
        draw=blue!65!black,
        inner sep=1.7pt
    },
    vright/.style={
        circle,
        draw=blue!65!black,
        fill=white,
        line width=0.8pt,
        inner sep=1.7pt
    },
    vstar/.style={
        star,
        star points=5,
        star point ratio=1.3,
        draw=black,
        fill=black,
        inner sep=1.2pt
    },
    link/.style={
        draw=gray!65!white,
        line width=0.9pt
    },
    linkr/.style={
        draw=red!50!black,
        line width=0.9pt
    },
    lab/.style={
        text=blue!25!black,
        font=\normalsize
    },
    score/.style={
        text=blue!25!black,
        font=\large
    },
    title/.style={
        text=black,
        font=\small
    },
    looparrow/.style={
        -{Stealth[length=1.2mm,width=1mm]},
        draw=gray!30!black,
        line width=0.4pt
    }
]


\newcommand{\CKMtriangle}[2]{%
\begin{scope}[shift={#1}, scale=#2]
    \coordinate (O) at (0,0);
    \coordinate (B) at (1,0);
    \coordinate (A) at (0.160,0.350);

    \pic[
    draw=red!70!black,
    line width=0.6pt,
    angle radius=1.5mm] {right angle=O--A--B};
    \draw[line width=0.6pt] (O) -- (B) -- (A) -- cycle;

    \fill (O) circle (0.01);
    \fill (B) circle (0.01);
    \fill (A) circle (0.01);

    \node[text=black,
        font=\fontsize{12pt}{17pt}\selectfont] at (0.35,0.1) {\textbf{Right}};
     \pic[draw=blue!29!gray!94!green!87!black,
        line width=0.6pt,
        angle radius=6mm,
        angle eccentricity=1.6,
    ] {angle=A--B--O};

\end{scope}
}

\CKMtriangle{(0,0)}{3};
\end{tikzpicture}}}
}{Title figure}\hspace{-0.4em}Theory of Flavor}

\author{Nima Arkani-Hamed$^{1}$}
\email{arkani@ias.edu}
\author{Carolina Figueiredo$^{2}$}
\email{cfigueiredo@fas.harvard.edu}
\author{Lawrence J. Hall$^{3,4}$}
\email{ljh@berkeley.edu}
\author{Claudio Andrea Manzari$^{1}$}
\email{manzari@ias.edu}
\affiliation{$^{1}$School of Natural Sciences, Institute for Advanced Study, Princeton, NJ 08540, USA \\
$^{2}$Society of Fellows, Harvard University, Cambridge, MA 02138, USA\\
$^{3}$Leinweber Institute for Theoretical Physics, Department of Physics, University of California, Berkeley, CA 94720, USA
\\
$^{4}$Theoretical Physics Group, Lawrence Berkeley National Laboratory, Berkeley, CA 94720, USA}

\begin{abstract}
A striking empirical observation about the CKM matrix is that the angles of the unitarity triangle $(\alpha, \beta, \gamma)$ are very close to $(\pi/2, \pi/8, 3 \pi/8) $, simple fractions of $\pi$ that are suggestive of an underlying theory linking flavor and spontaneous CP violation. However, relating this empirical observation to an underlying theory of flavor is challenging, since the unitarity triangle is a complicated function of the  Yukawa matrices. In this letter we present a simple picture for the Yukawas where this direct link is possible. We begin by parametrizing the ten-dimensional space of flavor data via ``nine-link textures",  full-rank $Y_{u,d}$ matrices with a total of nine non-zero entries, with a single CP violating phase.
Fitting the ten parameters of all such textures to the flavor data reveals a  wonderful surprise: the CP phases cluster tightly around multiples of $\pi/8$! This happens because the entries of $Y_{u,d}$
naturally define a ``Yukawa triangle", in most cases identical to the unitarity triangle at leading order in small flavor parameters. Most interestingly, these two  triangles are not the same beyond leading order, yielding precise predictions for $(\alpha, \beta, \gamma)$ with calculable deviation from $(\pi/2, \pi/8, 3 \pi/8)$, which can be decisively excluded or strongly confirmed by the next generation of experimental measurements of the angles. 
The 9-link textures are sparse and their determinants are naturally real, which taken together with spontaneous CP violation  can resolve the strong CP problem.

\end{abstract}

\maketitle

\noindent {\bf Seeking hidden patterns in flavor.}---The origin of fermion flavor remains one of the deepest open questions in particle physics. While the Standard Model (SM) successfully describes quark masses and mixing through two Yukawa matrices, it offers no explanation for the striking hierarchies among quark masses, the hierarchical structure of the Cabibbo-Kobayashi-Maskawa~\cite{Cabibbo:1963yz,Kobayashi:1973fv} (CKM) matrix, $V$, or the origin of CP violation. Ultimately, the Yukawa matrices are expected to emerge from a more fundamental theory in the ultraviolet.

For decades, theorists have stared at the  wealth of ever-more precisely measured flavor parameters, hoping to find simple and striking patterns that might provide vital clues to the origin of flavor in particular, and physics beyond the SM more generally -- perhaps in the same way the recognition of the Balmer formula for the spectrum of Hydrogen augured the coming of quantum mechanics. Some qualitative relationships are clearly visible: for instance the hierarchies seen in the masses are also reflected in the mixings via the rough relation $V_{i j} \sim \sqrt{m_i/m_j}$. But beyond this evidence for what we might call  ``generic" hierarchical structure, very precise relations (as hoped for in the Balmer formula analogy!) have been much harder to come by. Perhaps the best-known simple relation is a much more accurate version of the generic hierarchy relationship, which holds for the first two generations via $V_{us}\simeq \sqrt{m_d/m_s}$~\cite{Gatto:1968ss,Weinberg:1977hb,Fritzsch:1977za,Wilczek:1977uh,Wilczek:1978xi,Georgi:1979df}, and the connection to texture-zero Yukawa matrices~\cite{Fritzsch:1977vd,Wilczek:1977uh,Wilczek:1978xi,Fritzsch:1979zq,Georgi:1979df,Branco:1988iq,Dimopoulos:1991yz,Giudice:1992an,Hall:1993ni,Ramond:1993kv,Ibanez:1994ig,Barbieri:1996ww,Branco:1999nb,Roberts:2001zy}. While these have provided valuable guidance in the search for organizing principles behind flavor, it is fair to say no class of theories have emerged as a compelling ``canonical model of flavor", and so the search continues. 

\begin{figure}[H]
    \centering
    \input{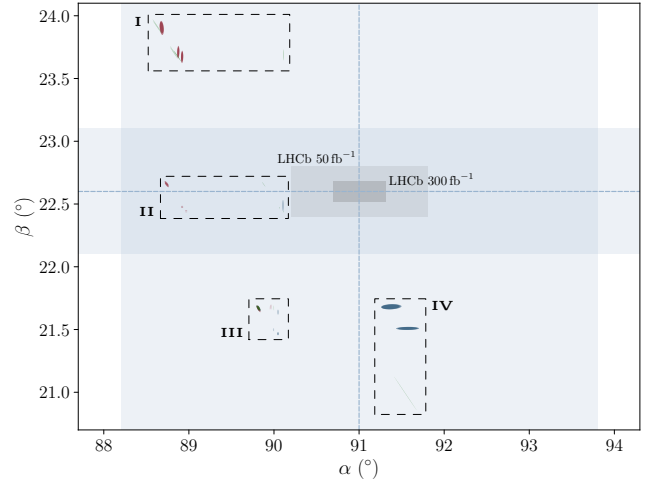}
    \vspace{-0.5cm}
    \caption{Each ellipse gives our most precise prediction for $(\alpha, \beta)$ from a particular 9-link texture at $2\sigma$. Current (blue) and projected (grey) $1\sigma$ experimental error bars are also shown \footnote{here the uncertainty in $\alpha$ is extrapolated from $(\beta,\gamma)$ using unitarity}.}
    \label{fig:predic}
\end{figure}

In this letter, we report on what we believe is a truly remarkable pattern hidden in flavor, with possibly dramatic implications for various aspects of physics beyond the Standard Model, which also makes extremely precise experimental predictions that will be decisively confirmed or excluded by future flavor experiments. As an illustration, a compendium of our most precise predictions is shown in Fig. \ref{fig:predic} \footnote{Note that the predictions are sufficiently precise that the corresponding ellipses in the figure are hard to see. A later figure will zoom in on the regions ${\rm I} \to {\rm IV}$ in the $(\alpha, \beta)$ plane populated by the predictions.}. 
\begin{figure*}[t]
    \centering    
    \input{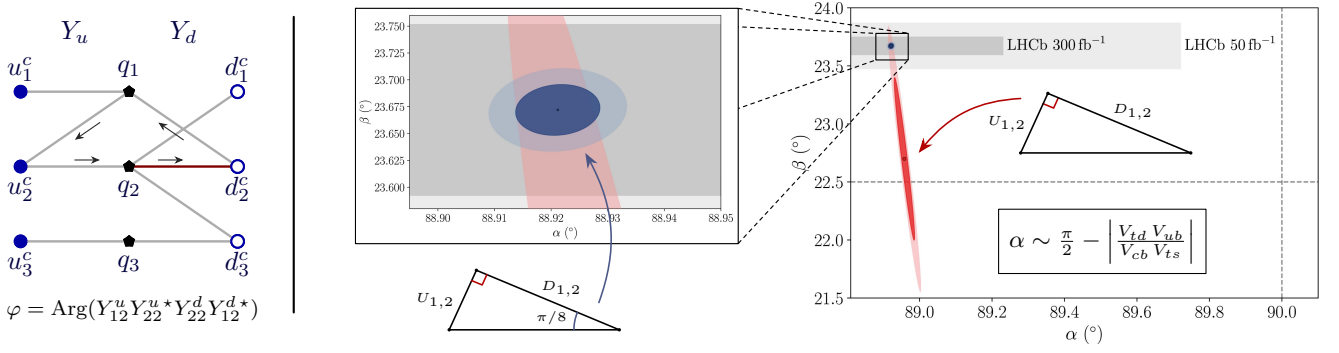}
    \caption{\textbf{(Left)} Example of a 9-link texture with a single rephasing-invariant, $\varphi$, associated with the closed loop; the phase is placed on entry $Y^d_{22}$ (marked in red). \textbf{(Right)} The ``Yukawa triangle" for this texture is formed by ratios $(U,D)_{1,2} = Y^{u,d}_{12}/Y^{u,d}_{22}$. Fixing this triangle to be exactly right predicts a calculable deviation for $\alpha$ away from $\pi/2$, and makes a narrow band of predictions in the $(\alpha, \beta)$ plane (red region on the right). Further fixing another angle  to be $\pi/8$, shrinks this prediction to a tiny region in the $(
    \alpha, \beta)$ plane, whose magnification is shown in blue. The spread in the predictions is due to the current $1$- and $2$-$\sigma$  experimental errors on the CKM matrix; the projected future experimental errors for  $(\alpha, \beta)$ are indicated by the grey boxes (centered around the blue ellipse for illustration). }
    \label{fig:IntroFigure}
\end{figure*}
\vspace{-0.5cm}
Our starting point is a different empirical observation about flavor, involving the angles of the famous ``unitarity triangle". Recall that this triangle is defined by the ``$d\,b$" entry of the unitarity relation for the CKM matrix $V_{ud}V_{ub}^{*} + V_{cd}V_{cb}^{*} + V_{td}V_{tb}^{*} = 0$; the angles of this triangle are then defined as 
\begin{equation}
 \scalebox{0.94}{$\alpha= \arg\!\left[\text{-}\frac{V_{td}V_{tb}^{*}}{V_{ud}V_{ub}^{*}}\right]\! , \,
  \beta= \arg\!\left[\text{-}\frac{V_{cd}V_{cb}^{*}}{V_{td}V_{tb}^{*}}\right] \!, \,
  \gamma= \arg\!\left[\text{-}\frac{V_{ud}V_{ub}^{*}}{V_{cd}V_{cb}^{*}}\right] \!.$} 
  \label{eq:angles_def}
\end{equation}
Decades of exquisite measurements at $B$ factories have determined these angles with remarkable accuracy, yielding~\cite{ParticleDataGroup:2026aaa}
\begin{equation}
  \beta = \bigl(22.6 \pm 0.5 \bigr)^{\circ}\,,\quad \gamma = \bigl(66.4^{+2.7}_{-2.8}\, \bigr)^{\circ}\,\label{eq:beta_and_gamma_exp}
\end{equation}
which when combined with the assumption of unitarity for the CKM matrix also determines
\begin{equation}
  \alpha = \bigl(91.0\pm 2.8\bigr)^{\circ}\,.
  \label{eq:alpha_UT}
\end{equation}
These precise measurements have in turn exposed a wonderful surprise: the angles of the unitarity triangle are not at all random, but lie  remarkably close to simple fractions of $\pi$,
\begin{equation}
\alpha \simeq \pi/2, \qquad
\beta \simeq \pi/8, \qquad
\gamma \simeq 3\pi/8,
\label{eq:npi/8}
\end{equation}
within current experimental precision.
 
These simple values are {\it especially} intriguing because discrete phases naturally arise in theories where CP is an exact symmetry of the Lagrangian and is broken spontaneously by the vacuum, which are in turn plausibly connected to theories for solving the strong CP problem.

It is then natural to ask: is the proximity of the measurements of $(\alpha, \beta, \gamma)$ to $(\pi/2, \pi/8, 3 \pi/8)$ merely an accident, or does it reflect a deeper structure connecting the physics of flavor and CP in the ultraviolet? A first indication that is encouraging for the pattern of angles ``not being an accident" is the nice fact that $(\alpha, \beta, \gamma)$ hardly run under the RG (see App. \ref{app:CKM_RGE}), so whatever pattern we see at the weak scale is just the same as what we would see at the ultraviolet scale where flavor is generated. 

But a moment's reflection also makes it clear that connecting a striking pattern in the unitarity triangle to a deeper theory for the fundamental Yukawa matrices themselves should be challenging. There are many simple theoretical mechanisms -- involving symmetries, anomalous dimensions, geography in extra dimensions, possibly combined with spontaneous CP violation -- that can impose simple patterns at the level of the Yukawa matrices, which are, after all, the fundamental parameters in the theory. By contrast, the unitarity triangle itself is {\it not} directly reflected in the Lagrangian, and not simply related to the Yukawa matrices. The sides of the triangle are highly ``processed" via diagonalizing the matrices, and are, in general,  complicated algebraic functions of the entries of the Yukawa matrices. Thus it appears unlikely that some sort of ``simplicity" at the level of the Yukawas could be directly connected to ``simplicity" at the level of the unitarity triangle, and vice-versa. 

This is the central surprise of our work: there {\it is} a picture for the Yukawa matrices, where the ``highly processed" unitarity triangle is in fact directly reflected in very simple properties of the matrices. 

We consider full-rank $Y_{u,d}$ matrices with a total of nine zeros and a single rephasing invariant -- these can be represented pictorially with ``link-diagrams'' (see Fig.~\ref{fig:IntroFigure}, left), with nodes representing the fields and links denoting the non-zero entries, which are placed such that there is a single closed loop, associated with the CP violating phase. We will see that in some cases the links in this closed loop themselves naturally define two sides of a triangle, the ``Yukawa triangle", which at leading order in small flavor parameters, is the same as the unitarity triangle! Thus simple assumptions about this ``Yukawa triangle" make predictions about the unitarity triangle. For example, we can demand that only one of the angles of this triangle is (depending on the texture) $(\pi/2,\pi/8,  3 \pi/8)$ and that leads to a prediction for the corresponding angle in the unitarity triangle. Or we can demand further that the angles of the Yukawa triangle are exactly $( \pi/2,\pi/8, 3 \pi/8)$ and this predicts the full unitarity triangle (see Fig.~\ref{fig:IntroFigure}, right).
But most interestingly and importantly, the correspondence between the ``Yukawa triangle" and the ``unitarity triangle" is not exact beyond leading order in small flavor parameters,  but has precisely calculable corrections. Thus, the demanding ``simplicity" in the angles of the Yukawa triangles, leads to {\it very precisely calculable deviations} from these special values in the actual unitarity triangle! 

We find this particularly exciting, because the next generation of flavor measurements at Belle II and LHCb are slated to improve the experimental errors on the angles by nearly an order of magnitude~\cite{Belle-II:2018jsg, LHCb:2018roe, ATLAS:2025lrr, LHCb:2021glh, Vagnoni:2025qfv}. In Fig.~\ref{fig:IntroFigure} we illustrate an example of the precise predictions from one texture, also indicating the projected future error bars -- making clear how future sensitivities will play a decisive role in either ruling out or confirming these predictions. As we will see in the rest of this letter, this conclusion extends to considering the predictions of all possible nine link textures, whose more systematic description we turn to now.\\

\noindent {\bf Parametrizing  flavor with 9-link textures.}--- Amusingly, we were led to our results not by directly thinking about connections between the unitarity triangle and the Yukawas, but rather by looking for an invariant way of parametrizing the flavor data. In this representation, the data naturally reveals the surprise that makes our subsequent theory for the unitarity angles possible. We believe this way of looking at flavor is of independent general interest and deserves further attention beyond the implications for the unitarity triangle. 

As we have recalled, the connection between flavor observables and the fundamental Yukawa matrices is highly non-unique. The $3 \times 3$ Yukawa matrices, $Y_{u,d}$, have $2 \times 3^2 \times 2 = 36$ real degrees of freedom, while there are only 10 flavor observables. Of course, all those $Y_{u,d}$ matrices that are related by the $U(3)^3 = U(3)_Q \times U(3)_{u^c} \times U(3)_{d^c}$ flavor symmetries are physically indistinguishable in the low-energy effective theory of the SM. Thus flavor space is really $36 - 3 \times 3^2 + 1 = 10$ dimensional, and the most physical way to parametrize the equivalence class of all matrices up to the action of $U(3)^3$ is to specify the 10 flavor parameters contained in the 6 eigenvalues, 3 angles, and 1 phase in the CKM matrix. While this parametrization of flavor is completely physical and invariant, it is not directly useful to the flavor model-builder, for whom the full 36-dimensional space is meaningful. This is because UV theories of flavor introduce more fields and new dynamics that break the $U(3)$ symmetries, so different forms of $Y_{u,d}$ suggest inequivalent UV origins for flavor. 
\begin{figure}[t]
    \centering
    \includegraphics[width=\linewidth]{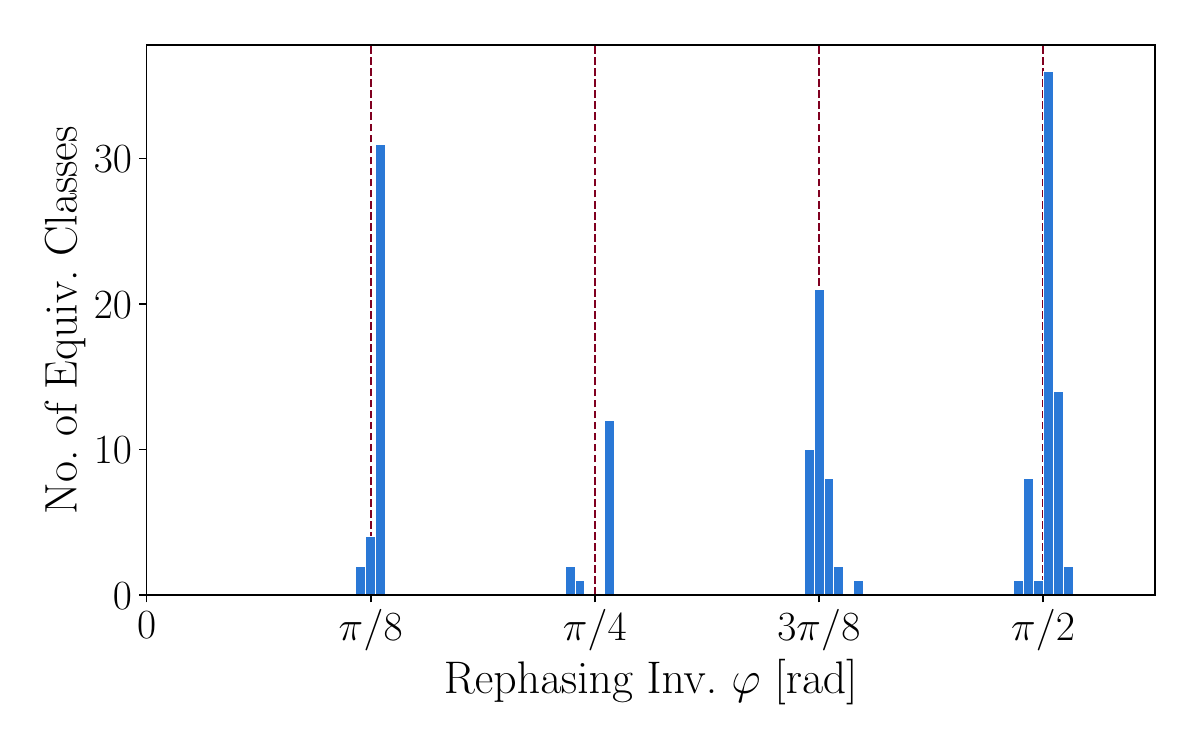}
    \caption{Histogram of rephasing invariant, $\varphi$ (up to complex-conjugation and $\pm \pi$), and respective number of permutation inequivalent fits (equivalence classes).}
    \vspace{-0.4cm}
    \label{fig:theta_histogram}
\end{figure}

It is then natural to ask: can we parametrize the flavor data directly in terms of quark Yukawa matrices $Y_u(l_i), Y_d(l_i)$ depending on only ten parameters $l_{i=1, \cdots 10}$, so that the 10 masses and mixings fix the $l_i$? 
The most natural way of doing this is to set to zero as many entries of $Y_{u,d}$ as needed to be left with the correct 10 d.o.f. -- $9$ entries and $1$ phase, which precisely leads to the 9-link textures described above. Since the resulting Yukawa matrices are fairly sparse, it is convenient to represent them via the link diagrams, as in Fig.~\ref{fig:IntroFigure}. By rephasing the fields (nodes), we can make every single link real and positive, except for one of the links in the closed loop -- which must carry the rephasing-invariant CP violating phase, $\varphi$. Of course there are many combinatorial choices for where to place the zeros, so there are many parametrizations of this sort, which are by design all equivalent under $U(3)^3$ rotations.  
Fitting the data to these 9 link textures gives us an alternative way of presenting the data of flavor in a way more transparently connected to the UV origin of flavor.

To cover the full parameter space, we perform a systematic scan over all $9$-link textures, $Y_{u,d}$. For each texture, we fit the $9$ real positive parameters and the CP phase, $\varphi$, to the $6$ quark masses and $4$ CKM parameters. Details of the scan and fitting procedure are given in SM. Among all textures, we find 156 different solutions, which are not related by trivial row/column permutations.  

We then did the obvious exercise of making a histogram of the fit value of $\varphi$ \footnote{throughout the text when we refer to $\varphi$ we are always refering to the phase we get in the rephasing invariant up to complex conjugation as well as adding/subtracting $\pi$.}; naively one would expect this histogram to give us a featureless, uninteresting distribution. Surprisingly, as shown in Fig.~\ref{fig:theta_histogram}, the result is completely different: the fit values of the rephasing-invariant cluster very tightly around a few values: $(\pi/8, \pi/4, \pi/2, 3 \pi/8)$! It was then immediate to recognize $(\pi/8, \pi/2, 3 \pi/8)$ as the angles of the unitarity triangle, strongly suggesting that the triangle was directly reflected in the textures! The  histogram also shows an interesting peak at $ \varphi \sim \pi/4$ which, as we explain in App.~\ref{app:Pi4}, reflects yet \textit{another} pattern in flavor, not directly connected to the unitarity triangle! 
\begin{figure*}[p!]
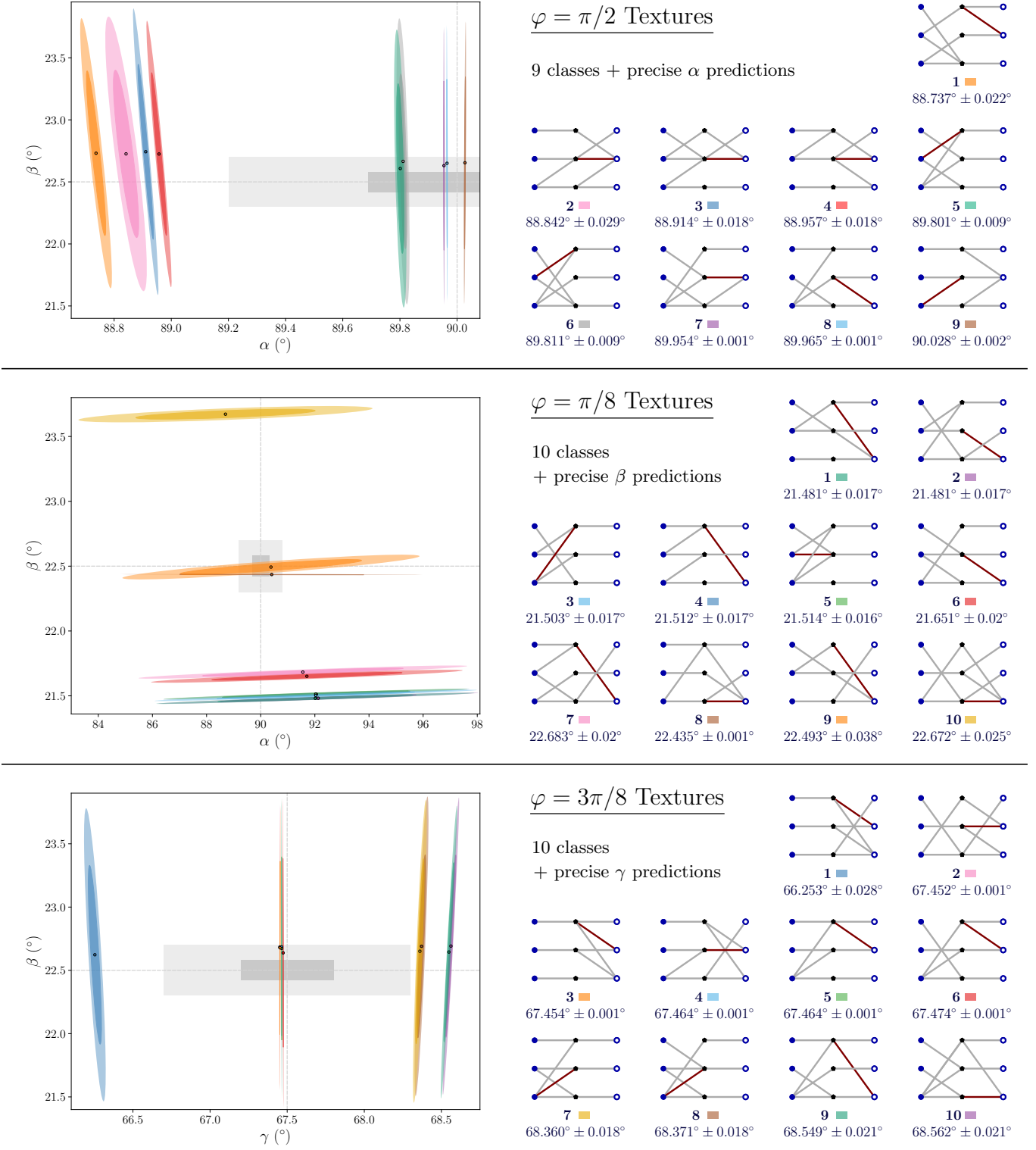

        \centering

        \input{Figures/AllClasses_new_alpha}

        \par\vspace{-0.8em}
        \rule{\linewidth}{0.6pt}
        \par\vspace{0.7em}

        \input{Figures/AllClasses_new_beta}

        \par\vspace{-0.8em}
        \rule{\linewidth}{0.6pt}
        \par\vspace{0.7em}

        \input{Figures/AllClasses_new_gamma_2}

        \caption{One and two sigma regions for the prediction of the angles of the CKM unitarity triangle, obtained by only fixing the rephasing-invariant phase, $\varphi$.  \textbf{(Top Panel)} The nine equivalence classes of textures obtained by fixing $\varphi = \pi/2$ and their predictions plotted in the $(\alpha, \beta)$ plane; \textbf{(Middle Panel)} The ten equivalence classes for $\varphi=\pi/8$ and their predictions in the $(\alpha,
        \beta)$ plane; \textbf{(Bottom Panel)} The ten (non-accidental) equivalence classes for $\varphi = 3 \pi/8$ and their predictions in the $(\beta, \gamma)$ plane. The spreads are entirely due to current experimental uncertainties on the flavor observables used in the fit. In theories where a given CKM angle is predicted to coincide with $\varphi$ at leading order, the spread in that angle is considerably smaller than in the other two, leading to elongated 1- and 2-sigma regions. The gray boxes (centered at $(\pi/2,\pi/8,3\pi/8)$) denote the expected LHCb sensitivities for early 2030s and 2040s.}
        \label{fig:PhaseClasses}
\end{figure*}
\\

\noindent {\bf Fixing the Yukawa rephasing invariant.}--- 
Taking the histogram seriously, it is natural to wonder whether we can still find a fit if we {\it fix} the rephasing invariant to be $ \{\pi/2,\pi/8,3\pi/8\}$ -- which would let us predict $\{ \alpha, \beta, \gamma\}$. Repeating the scan with fixed $\varphi$, we find, up to row/column permutations, 29 $(\pi/2)$, 35 $(\pi/8)$, and 35 $(3\pi/8)$ 9-link textures yielding fits with $\chi^2$ values below the $3\, \sigma$ confidence-level threshold. We show the link-diagrams for all these classes in App.~\ref{app:Fit}. In the full scan, for a given texture we find different fits, corresponding to different local minima. Most of these, however, do not exhibit a ``hierarchical structure'' --  the left- and/or right-diagonalizing matrices contain large mixing angles and are therefore far from diagonal. Interestingly, within each permutation orbit there is at most one hierarchical fit, by which we mean that all diagonal entries of both the left- and right-diagonalizing matrices are $\geq 0.9$. For orbits in which this criterion cannot be fulfilled, there is nevertheless always one fit that is hierarchical on the left, but not on the right. In the rest of the text, we always choose the most hierarchical texture as a representative element of each equivalence class. 

In addition to more transparently reflecting the data, having a hierarchical structure is practically useful as it allows for a controlled perturbative diagonalization, from which we can derive analytic approximations for the quark masses and the CKM matrix. Assuming this hierarchical structure of $Y_{u,d}$, in App.\ref{app:U(1)9}, we give a general analytic derivation of the leading order predictions for 
$\alpha, \beta$ or $\gamma$. We find that if $Y^{u,d}_{ij}$ have phases that are multiples of $\pi/8$, this translates into $\alpha, \beta$ or $\gamma$ being a multiple of $\pi/8$ {\it only if there are texture zeros} \footnote{Here, by ``vanishing'' we mean that the size of the entries are less than a percent of the ``natural'' hierarchical entries for the Yukawa matrix.}! Hierarchical matrices can be approximately diagonalized by successive $2\times2$ unitary transformations, and the texture zeros enforce conditions on these transformations that are necessary for nailing the special values of the angles. 
We also show analytically that all realistic 9-link textures have $\varphi$ close to a multiple  $(1,3,4)\pi/8$, $\varphi$ is approximately $(\beta, \gamma, \alpha)$, respectively.

Note, however, that once we fix $\varphi$, it is no longer guaranteed we can map any textures into each other using $U(3)^3$ rotations -- since these might not preserve $\varphi$. Modding-out by those rotations that preserve the phase, defines the equivalence classes which lead to genuinely different physical predictions: $9$ classes for $\varphi=\pi/2$;  $10$  for $\varphi=\pi/8$ and $13$ for $\varphi=3\pi/8$. Applying perturbative diagonalization each class, we find as expected that in most cases, one of the CKM angles $(\alpha,\beta,\gamma)$, at leading order, is determined by $\varphi$! The only exception happens for three classes for $\varphi=3\pi/8$, where an accident between Yukawa entries ensures $\gamma \sim 3\pi/8$, as discussed in App.~\ref{app:Fit}. Removing these three cases, we are left with $(9,10,10)$ classes for $ \varphi = \{\pi/2,\pi/8,3\pi/8\}$, respectively. 

Since these theories contain only nine independent real parameters, which are fit to the ten flavor observables, each class predicts one non-trivial relation among the CKM observables, which in this setting are the precise deviations from the exact values of $\alpha \sim \pi/2$, $\beta \sim \pi/8$ and $\gamma \sim 3 \pi/8$. In App.~\ref{app:Examples}, we show some examples of how the perturbative diagonalization of a 9-link texture lets us derive not only the leading contribution to each angle but also the precise sub-leading corrections, which can be fully recast in terms of the observables. In Fig.~\ref{fig:PhaseClasses} we present the different theories (9,10,10) and their respective precise predictions for $(\alpha,\beta,\gamma)$. 
\begin{figure*}[t]
    \centering
    \input{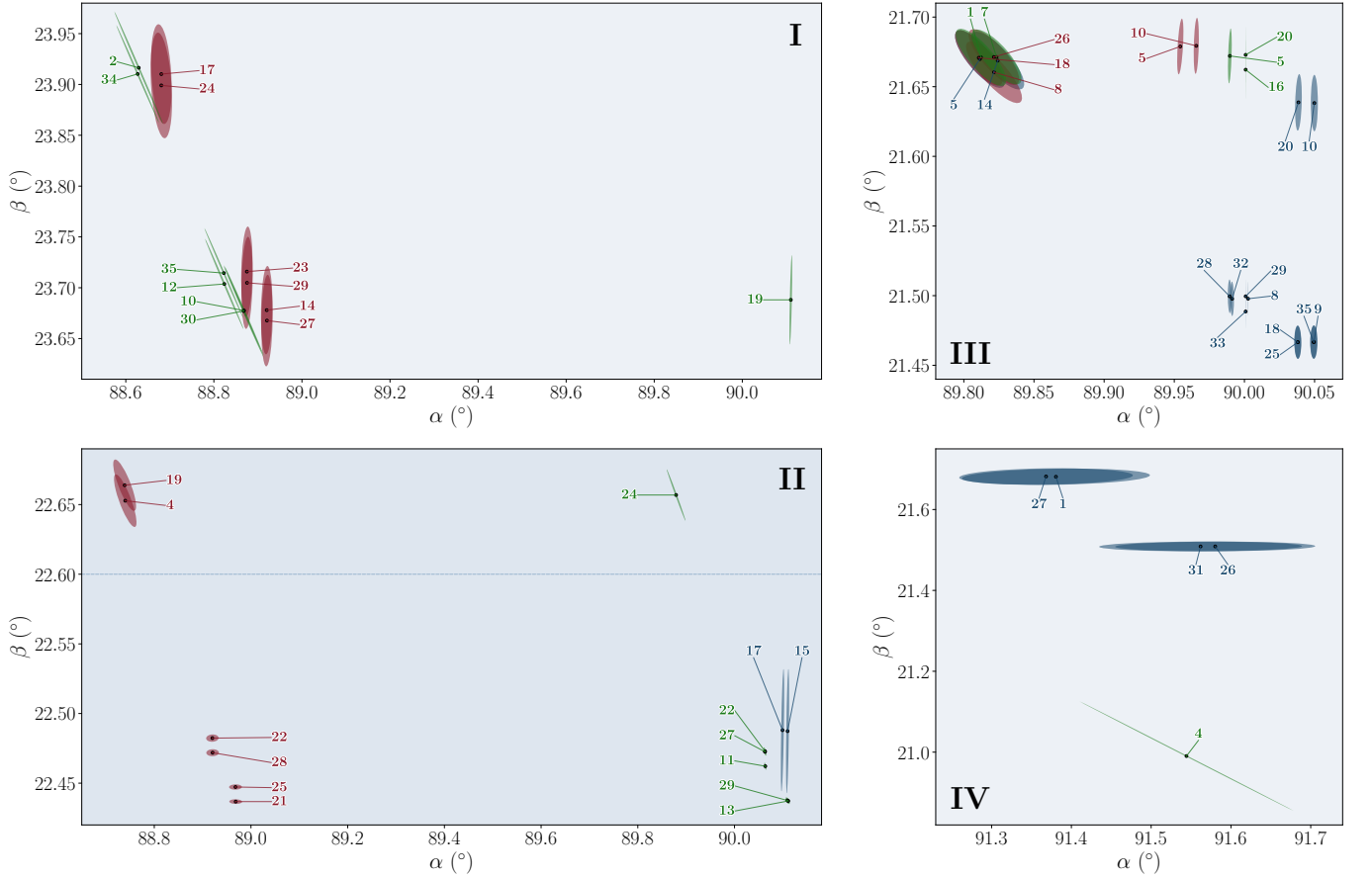}
    \caption{Four zoom-ins from Fig. \ref{fig:predic}. Two-sigma prediction regions in the $\alpha$--$\beta$ plane obtained by fixing the entire unitarity triangle at leading order through the Yukawa triangle. Equivalence classes with $R_{\alpha}=i/\tan(\pi/8)$ are shown in \textbf{red}, those with $R_{\beta}=e^{i\pi/8}/\cos(\pi/8)$ in \textbf{blue}, and those with $R_{\gamma}=e^{i3\pi/8}\cos(3\pi/8)$ in \textbf{green}. Each region is labeled by the corresponding texture number in Figs.~\ref{fig:29textures_pi2}--\ref{fig:35textures_3pi8}.}
    \vspace{-0.2cm}
    \label{fig:ZoomIns}
\end{figure*}

Our precise predictions arise from having as many zeros in $Y_{u,d}$ as possible, but, obviously, these predictions can still hold to high accuracy as we deform away from 9-link textures. The analysis of 
App.~\ref{app:U(1)9} shows that texture zeros are still necessary for this to happen: there are many ways that textures and spontaneous CP violation can lead to one angle of the unitarity triangle being a $(1,3,4)$ multiple of $\pi/8$ at leading order; but at least four zeros (and of course up to nine) appear in all patterns. 
Apart from this analytic understanding,
a preliminary numerical study of these perturbations is also presented, confirming the many zeros that are needed to predict the angles close to the special fractions of $\pi$. We leave a more systematic exploration of the deviations from 9-link textures to future work.\\ 

\noindent {\bf Fixing the Yukawa triangle.}--- So far we have focused on how we can control one of the angles of the unitarity triangle via the rephasing-invariant. For example, for $\alpha$ the relevant ratio is
\begin{equation}
 -\frac{V_{td}V_{tb}^{*}}{V_{ud}V_{ub}^{*}} \equiv |R_{\alpha}| e^{i \alpha},
\end{equation}
and in the $\pi/2$-textures we find that the leading order expansion of these CKM entries are precisely such that the argument of the ratio is given by $\varphi$. Knowing that, at leading order, the unitarity triangle is a right triangle whose two orthogonal sides are  $V_{td}V_{tb}^{*}$ and $V_{ud}V_{ub}^{*}$, it follows that $\beta$ is such that $\tan^{-1}(\beta) = |R_{\alpha}|$. By perturbatively diagonalizing the different textures, we find that in most cases not only $\alpha \sim \varphi$, but also $|R_{\alpha}|$ is precisely the rephasing-invariant monomial! For example, in class $4$ of the $\pi/2$ theories  -- the class presented in Fig.\ref{fig:IntroFigure} -- we find that, at leading order, 
\begin{equation}
\frac{V_{td} V^*_{tb}}{V_{ud} V_{ub}^*} \sim \frac{D_{12}}{U_{12}}, \quad \text{with } \, D_{12} = \frac{Y^d_{12}}{Y^d_{22}}, \, \, U_{12}= \frac{Y^u_{12}}{Y^{u}_{22}} \, .
  \label{eq:LeadOrdAlpha}
\end{equation}
Given that $|U_{12}|$ and $|D_{12}|$ are of the same order, it is then meaningful to think of the complex numbers, $U_{12}$ and $D_{12}$, as defining the sides of a triangle -- the \textit{Yukawa triangle} (see Fig. \ref{fig:IntroFigure}). By fixing the argument of the ratio to be $-\pi/2$ we obtain a right triangle, but we can fully fix the triangle (up to rescalings) by further demanding 
\begin{equation}
   \frac{U_{12}}{D_{12}}  = i\tan{\frac{\pi}{8}}\,.
   \label{eq:fixedtr}
\end{equation}
By construction we have that at leading order the unitarity triangle is exactly the Yukawa triangle, with angles $(\pi/2,\pi/8,3\pi/8)$; and the subleading corrections are totally calculable!

Looking at the results of perturbative diagonalization for the different classes in Fig.$\,$\ref{fig:PhaseClasses}, we find that the simplicity of $R_\alpha$ described for class $4$, is quite general, and generalizes to $R_{\beta}$ and $R_{\gamma}$ in the $\pi/8$ and $3\pi/8$ theories -- in most cases the ratio is precisely the monomial in the rephasing invariant (see Tab.~\ref{tab:LeadingOrderAngles})! 
It is then natural to ask whether we can still fit the data by not only fixing $\varphi$, but the \textit{full} Yukawa triangle. In the $\pi/2$ case this corresponds to fixing the leading $|R_{\alpha}| = 1/\tan(\pi/8)$; for $\pi/8$ and $3\pi/8$ instead we have $|R_{\beta}| = 1/\cos(\pi/8)$ and $|R_{\gamma}| = \cos(3\pi/8)$. In such a fit, we predict precisely two angles of the unitarity triangle, which assuming unitarity  determines the full triangle. 
To do this analysis we go back to the $(29,35,35)$ hierarchical representatives of the permutation inequivalent textures (represented in Figs.  \ref{fig:29textures_pi2}-\ref{fig:35textures_3pi8}); and fix not only $\varphi$, but also the respective rephasing invariant monomial to $\{\tan{\pi/8},\cos{\pi/8},\cos{3\pi/8}\}$, respectively. Fitting to the data we find that $(23,27,33)$ yield fits with $\chi^2$ values below the $3\, \sigma$ confidence-level threshold. (see details in SM). As previously, to account for the textures which give physically different predictions we need to further mod-out by the subset of $U(3)^3$ rotations which preserves not only the phase but the full rephasing invariant. In doing so we are left with $(17,19,19)$ different textures which yield extremely precise predictions for $(\alpha,\beta,\gamma)$ as shown already in Fig.~\ref{fig:predic}. In Fig.~\ref{fig:ZoomIns}, we present the zoom-ins on the different clusters of predictions, showing how small the $2$-sigma ellipses are once we impose the constraint on the full Yukawa triangle.
\\

\noindent {\bf Spontaneous CP violation and Strong CP.}---As already remarked above, the simple fractions of $\pi$ seen in the unitarity triangle are suggestive of an underlying theory of spontaneous CP violation. Realizing these angles directly from simple phases in the Yukawa matrices cries out for this interpretation with much greater force.

There are several simple pictures for correlating the physics of flavor generation and spontaneous CP violation that can guarantee the entries in the Yukawa matrices have simple phases giving rise to our predictions. For instance, CP could be spontaneously broken by the vev of a complex scalar field, $\chi$, whose potential has a $\mathbb{Z}_4$ symmetry, giving rise to vevs that are purely real or purely imaginary -- trivially guaranteeing that the phase in the ``Yukawa triangle" is exactly $\pi/2$. Similar ideas can be used to determine the full shape of the ``Yukawa triangle" to match the $(\pi/2,\pi/8, 3 \pi/8)$ triangle. For example, a combination of discrete and flavor symmetries can guarantee relations of the form $Y^{u,d}_{12} = (\chi_1^{u,d} + \chi_2^{u,d})Y^{u,d}_{22}$ where the $\chi$'s get vevs respecting a $\mathbb{Z}_8$ global symmetry, thereby having equal magnitudes but phases that are multiples of $\pi/4$. This can yield $U_{12}/D_{12} =  (Y^{u}_{12}/Y^u_{2,2})/(Y^d_{12}/Y^d_{22}) = i (1 - e^{i \pi/4})/( 1 + e^{i \pi/4}) = i {\rm tan}(\pi/8)$. We defer a discussion of these theoretical possibilities to future work, as these are not directly relevant for the UV detail-independent, precise experimental predictions we wish to focus on in this letter. 

But there is a final obvious point worth making in connection with spontaneous CP violation. As we have seen, the 9-link textures give $Y_{u,d}$ matrices that are as sparse as possible to fit the data. These matrices have so many zeros, that with no special contrivance, in most cases, the simple phases can be placed in links such that their determinants are automatically real. The link-diagrams make it especially easy to see this: each term in the determinant is given by a \textit{perfect matching} of the graph -- a collection of edges touching all nodes exactly once. For example, in the texture in Fig.$\,$\ref{fig:IntroFigure}, there is a single term in Det($Y_u Y_d$), and therefore a single perfect matching for both the $Y_u$ and $Y_d$ matrices, which is 
\begin{equation}
    \begin{gathered}
        \begin{tikzpicture}[
    scale=1.0,
    line cap=round,
    line join=round,
    vleft/.style={
        circle,
        fill=blue!65!black,
        draw=blue!65!black,
        inner sep=1.3pt
    },
    vright/.style={
        circle,
        draw=blue!65!black,
        fill=white,
        line width=0.8pt,
        inner sep=1.3pt
    },
    vstar/.style={
        star,
        star points=5,
        star point ratio=1.3,
        draw=black,
        fill=black,
        inner sep=1.2pt
    },
    link/.style={
        draw=gray!65!white,
        line width=0.9pt
    },
    links/.style={
        draw=red!50!black,
        line width=0.9pt
    },
    lab/.style={
        text=blue!25!black,
        font=\footnotesize
    },
    score/.style={
        text=blue!25!black,
        font=\small
    },
    title/.style={
        text=blue!65!black,
        font=\large
    },
    matrixnode/.style={
    font=\scriptsize,
    text=black
    },
    highlight/.style={
    draw=blue!50!black,
    line width=3pt,
    opacity=0.25
    }
]

\newcommand{\graphnodes}{
    \node[vleft]  (u1) at (0, 1) {};
    \node[vleft]  (u2) at (0, 0) {};
    \node[vleft]  (u3) at (0,-1) {};

    \node[vstar]  (q1) at (1.45, 1) {};
    \node[vstar]  (q2) at (1.45, 0) {};
    \node[vstar]  (q3) at (1.45,-1) {};

    \node[vright] (d1) at (2.90, 1) {};
    \node[vright] (d2) at (2.90, 0) {};
    \node[vright] (d3) at (2.90,-1) {};
}

\newcommand{\graphlabels}{
    \node[lab] at (0, 0.75) {$u^c_1$};
    \node[lab] at (0, -0.25) {$u^c_2$};
    \node[lab] at (0, -0.75) {$u^c_3$};
    \node[lab] at (1.45, 0.75)
    {$q_1$};
    \node[lab] at (1.45, -0.25)
    {$q_2$};
    \node[lab] at (1.45, -0.75)
    {$q_3$};
    \node[lab] at (2.90, 0.73) {$d^c_1$};
    \node[lab] at (2.90, -0.27) {$d^c_2$};
    \node[lab] at (2.90, -0.75) {$d^c_3$};
}

\begin{scope}[shift={(0.18,0)},scale=0.83]

    \graphnodes
    \graphlabels
    \draw[link] (u1) -- (q1);
    \draw[link] (u2) -- (q2);
    \draw[link] (u3) -- (q3);
    \draw[link] (u2) -- (q1);

    \draw[link] (q2) -- (d1);
    \draw[link] (q1) -- (d2);
    \draw[link] (q2) -- (d2);
    \draw[link] (q2) -- (d3);
    \draw[link] (q3) -- (d3);
    \draw[links] (q2) -- (d2);

     \draw[highlight] (u2) -- (q2);
     \draw[highlight] (u1) -- (q1);
     \draw[highlight] (u3) -- (q3);

     \draw[highlight] (d2) -- (q1);
     \draw[highlight] (d1) -- (q2);
     \draw[highlight] (d3) -- (q3);

    \graphnodes

\end{scope}

\end{tikzpicture}
    \end{gathered} \, \,  \Rightarrow  \, \,\begin{aligned} 
    &\text{Det}(Y_u) = Y^u_{11}Y^u_{22}Y^u_{33} \\
    &\text{Det}(Y_d) = Y^d_{12}Y^d_{21}Y^d_{33} \\
    \end{aligned} \, .
\end{equation}
So if we place the phase in $Y^u_{12}$ or $Y^d_{22}$, the determinant will remain real. As it turns out, almost all the $9$-link textures have exactly this feature -- there is a single perfect matching in $Y_u$ and $Y_d$, and so one can always place the phase somewhere in the loop such that $\arg(\det(Y_u Y_d)) = 0$. Out of the $(29+35+35)$ textures, only $5$ fail this criterion (marked with $(*)$ in Figs.\ref{fig:29textures_pi2}-\ref{fig:35textures_3pi8}) --  all $5$ are diagonal in one of the sectors. Thus, for the vast majority of textures, having $\arg(\det(Y_u Y_d)) = 0$ eliminates the Standard Model contributions to the QCD $\theta$ angle, and suggests a full solution to the strong CP problem in the UV, with no appeal to axions. \\

\noindent {\bf Relation to previous work.}--- There is a vast literature on texture zero models for flavor, covering many patterns of zeros \cite{Fritzsch:1977vd,Wilczek:1977uh,Wilczek:1978xi,Fritzsch:1979zq,Georgi:1979df,Branco:1988iq,Dimopoulos:1991yz,Giudice:1992an,Ramond:1993kv,Hall:1993ni,Ibanez:1994ig,Barbieri:1996ww,Branco:1999nb,Kuo:1999dt,Chiu:2000gw,Roberts:2001zy,Xing:2015sva}, including those that emphasize 9-link textures as a canonical way of parameterizing flavor \cite{Branco:2004ya,Ludl:2015lta,Giraldo:2015ffa,Emmanuel-Costa:2016gdp}. As we have emphasized, the central result motivating our work is an empirical surprise which, to our knowledge, had not been noted in previous literature -- the striking clustering of the phase $\varphi$ around multiples of $\pi/8$ in scanning over all possible 9-link textures. 
There is also a related large literature on special choices of zeros and phases that can predict aspects of the unitarity triangle \cite{Masina:2006ad,Masina:2006pe,Xing:2009eg,Antusch:2009hq,Harrison:2009bb,Harrison:2009bz,Antusch:2011sx,Antusch:2013rla,Mimura:2018sbc}. The relation between $\alpha \simeq \pi/2$, theories of spontaneous CP violation and the strong CP problem has also been emphasized~\cite{Antusch:2013rla}.
An especially interesting recent paper~\cite{Harrison:2025rkp} is also motivated by the same $(\alpha \sim \pi/2, \beta \sim  \pi/8, \gamma \sim 3\pi/8)$ observation, which is used as motivation to consider a special pattern of Hermitian Yukawa matrices with two zeros each, and a constraint as in~\eqref{eq:fixedtr} --  which yields a precise prediction for $(\alpha, \beta, \gamma)$ that differs from all of ours. More generally, much of the previous literature focuses on specific sets of textures (with various theoretical motivations), whereas we have focused on characterizing precise predictions for $(\alpha, \beta, \gamma)$ coming from all possible 9-link textures. It is striking that even taking all possible 9-link textures together covers only a minuscule fraction of the experimentally allowed space in $(\alpha, \beta, \gamma)$. We feel that these tiny islands furnish canonical ``lampposts" in $(\alpha, \beta, \gamma)$ space, which we can also perturb away from in a systematic way.\\

\noindent {\bf Outlook.}--- The current experimental determination of the unitarity triangle has put us in an extremely interesting position vis-a-vis the question of deeper structures underlying flavor. The measurements are suggestive of a striking pattern, but the error bars are still large enough to be consistent with nothing more than a coincidence. The order of magnitude reduction in error bars promised by the next generation of measurements thus carry enormous significance.

We find it fascinating that looking at the experimental flavor data through the lens of 9-link textures reveals a striking simplicity. Despite the great variety of these textures, the single irremovable phase is {\it always determined to be close to a multiple of $\pi/8$}, strongly suggesting spontaneous CP violation and a solution to the strong CP problem, all in one fell swoop.
This simple picture for flavor populates the allowed region in $(\alpha, \beta, \gamma)$ space with miniscule islands having clean and direct theoretical interpretations. It is even more exciting that the extremely precise predictions for $(\alpha, \beta, \gamma)$ resulting from this picture can be decisively excluded or strongly confirmed by the planned level of precision of the next generation of measurements of these angles.

While we have focused on the sharpest predictions for the angles arising from 9-link textures, as  we have mentioned,  high-accuracy predictions for the angles with a few percent of their special values also hold when deforming away from nine-link textures, though still demanding four or more texture zeros. It will be very interesting to study these deformations more systematically, together with possible UV origins for textures leading to  sharp predictions for $(\alpha, \beta, \gamma)$.

Indeed there are doubtless many more interesting theories of flavor that can give rise to other islands in $(\alpha, \beta, \gamma)$ space. This state of affairs offers an important and exciting challenge for theorists to understand precise predictions for the unitarity triangle in the coming years, as we await the verdicts from experiment, and also a further inducement to experimentalists to try and make the most accurate possible measurements of these crucial observables. \\ 

\noindent {\bf Acknowledgements.}--- The work of N.A.H. is supported by the DOE (Grant No. DE-SC0009988), the Simons Collaboration on Celestial Holography, the European Union (ERC, UNIVERSE PLUS, 101118787), and the Carl B. Feinberg cross-disciplinary program in innovation at the IAS. C.F. is supported by the Society of
Fellows at Harvard University. The work of L.J.H. was supported by the NSF grant PHY-2515115 and the Office of High Energy Physics of the U.S. Department of
Energy under contract DE-AC02-05CH11231. The work of C.A.M. is supported by the U.S. Department of Energy (DE-SC0009988) and the Sivian Fund. Our numerical analyses were carried with the help of Claude Code. Views and opinions expressed are those of the author(s) only and do not necessarily reflect those of the European Union or the European Research Council Executive Agency. Neither the European Union nor the granting authority can be held responsible for them. 

\let\oldaddcontentsline\addcontentsline
\renewcommand{\addcontentsline}[3]{}

\bibliography{biblio}

\clearpage

\let\addcontentsline\oldaddcontentsline

\clearpage

\onecolumngrid
\begin{center}
  \textbf{\large Supplementary Material for The Very Nearly  
\texorpdfstring{\raisebox{-0.52em}{%
  \resizebox{3cm}{!}{\begin{tikzpicture}[scale=0.9,
    line cap=round,
    line join=round,
    vleft/.style={
        circle,
        fill=blue!65!black,
        draw=blue!65!black,
        inner sep=1.7pt
    },
    vright/.style={
        circle,
        draw=blue!65!black,
        fill=white,
        line width=0.8pt,
        inner sep=1.7pt
    },
    vstar/.style={
        star,
        star points=5,
        star point ratio=1.3,
        draw=black,
        fill=black,
        inner sep=1.2pt
    },
    link/.style={
        draw=gray!65!white,
        line width=0.9pt
    },
    linkr/.style={
        draw=red!50!black,
        line width=0.9pt
    },
    lab/.style={
        text=blue!25!black,
        font=\normalsize
    },
    score/.style={
        text=blue!25!black,
        font=\large
    },
    title/.style={
        text=black,
        font=\small
    },
    looparrow/.style={
        -{Stealth[length=1.2mm,width=1mm]},
        draw=gray!30!black,
        line width=0.4pt
    }
]


\newcommand{\CKMtriangle}[2]{%
\begin{scope}[shift={#1}, scale=#2]
    \coordinate (O) at (0,0);
    \coordinate (B) at (1,0);
    \coordinate (A) at (0.160,0.350);

    \pic[
    draw=red!70!black,
    line width=0.6pt,
    angle radius=1.5mm] {right angle=O--A--B};
    \draw[line width=0.6pt] (O) -- (B) -- (A) -- cycle;

    \fill (O) circle (0.01);
    \fill (B) circle (0.01);
    \fill (A) circle (0.01);

    \node[text=black,
        font=\fontsize{12pt}{17pt}\selectfont] at (0.35,0.1) {\textbf{Right}};
     \pic[draw=blue!29!gray!94!green!87!black,
        line width=0.6pt,
        angle radius=6mm,
        angle eccentricity=1.6,
    ] {angle=A--B--O};

\end{scope}
}

\CKMtriangle{(0,0)}{3};
\end{tikzpicture}}}
}{Title figure}\hspace{-0.4em}Theory of Flavor
  }\\[.2cm]
  \vspace{0.05in}
  {Nima Arkani-Hamed, Carolina Figueiredo, Lawrence J. Hall and Claudio Andrea Manzari}
\end{center}

\twocolumngrid

\setcounter{equation}{0}
\setcounter{figure}{0}
\setcounter{table}{0}
\setcounter{section}{0}
\setcounter{page}{1}
\makeatletter
\renewcommand{\theequation}{S\arabic{equation}}
\renewcommand{\thefigure}{S\arabic{figure}}
\renewcommand{\thetable}{S\arabic{table}}

\setcounter{secnumdepth}{2}
\renewcommand{\thesection}{\Roman{section}}
\renewcommand{\thesubsection}{\thesection.\alph{subsection}}
\renewcommand{\p@subsection}{}

\onecolumngrid

\startcontents[sections]
\tableofcontents

\section{The CKM Matrix and the Unitarity Triangle}

In the Standard Model (SM) the quark masses and flavor mixing arise from the Yukawa sector. After electroweak symmetry breaking by the Higgs vacuum expectation value $v$, the quark mass matrices are $M_u = v\,Y_u/\sqrt{2}$, $M_d = v\,Y_d/\sqrt{2}$, where $Y_{u,d}$ are the $3\times 3$ complex Yukawa matrices in generation space. Physical (mass) eigenstates are obtained by independent bi-unitary rotations,
\begin{equation}
  \hat{M}_u = V^u_L\,M_u\,V^{u\, \dagger}_R , \qquad
  \hat{M}_d = V^d_L\,M_d\,V^{d\, \dagger}_R ,
  \label{eq:biunitary}
\end{equation}
where $\hat{M}_{u} = \mathrm{diag}(m_u,m_c,m_t)$ and
$\hat{M}_{d} = \mathrm{diag}(m_d,m_s,m_b)$ are real and positive, and $V^u_L, V^u_R, V^d_L, V^d_R\in U(3)$. In the SM, only the left-handed rotations are physical as they enter in the charged-current interactions through the coupling of the $W$ boson and combine into the observable CKM matrix~\cite{Cabibbo:1963yz,Kobayashi:1973fv}
\begin{equation}
  V \equiv V^u_L\,V^{d\, \dagger}_L .
  \label{eq:CKM_def}
\end{equation}
The matrix $V$ is unitary by construction, $V^{\dagger}V=\mathbf{1}$, and enters the charged-current Lagrangian as
\begin{equation}
  \mathcal{L}_{\rm cc}
  = \frac{g}{\sqrt{2}}\,W_\mu^{+}\,
    \bar{u}_{L,i}\,\gamma^\mu\,V_{ij}\,d_{L,j}
    + \mathrm{h.c.}
  \label{eq:cc_lag}
\end{equation}
A $3\times 3$ unitary matrix has nine real parameters; five
independent quark-field rephasing freedoms remove five phases, leaving four physical parameters: three real mixing angles and one CP-violating phase $\delta$. 
The unique rephasing-invariant measure of CP violation is the Jarlskog invariant~\cite{Jarlskog:1985ht}, $J$, 
\begin{equation}
 J\sum_{m,n} \epsilon_{ijm}\epsilon_{kln} = \mathrm{Im}\!\left(V_{ij}V_{kl}V_{il}^{*}V_{kj}^{*}\right) ,
 \label{eq:jarlskog}
\end{equation}
which vanishes if and only if CP is conserved in the quark sector. Experimentally, $J \simeq 3.12\times 10^{-5}$~\cite{PDG2024}, and therefore CP is violated by the weak interactions.\\

The unitarity of $V$ implies six off-diagonal orthogonality conditions.
The relation involving the first and third columns,
\begin{equation}
  V_{ud}V_{ub}^{*} + V_{cd}V_{cb}^{*} + V_{td}V_{tb}^{*} = 0 ,
  \label{eq:UT_relation}
\end{equation}
defines a triangle in the complex plane, one of six such triangles
implied by unitarity.  All six have the same area $J/2$, and
therefore encode identical information about CP violation. However, measuring this area experimentally requires $\mathcal{O}(1)$ interior angles; for the other five triangles one side is much shorter than the remaining two, and so the triangle is very squashed and the associated CP asymmetries in meson decays are suppressed.
The triangle of
Eq.~\eqref{eq:UT_relation} -- the \emph{unitarity triangle} (UT) -- has all three sides of comparable magnitude~\cite{PDG2024}, $\left|V_{ud}V_{ub}^{*}\right| \simeq 3.6\times 10^{-3}$, $\left|V_{cd}V_{cb}^{*}\right| \simeq 9.4\times 10^{-3}$, and so all three angles are $\mathcal{O}(1)$ and the area $J/2$ is directly accessible through $B$-meson decay asymmetries. Dividing Eq.~\eqref{eq:UT_relation} by $V_{cd}V_{cb}^{*}$ and normalizing the base to unity, the three interior angles are defined as~\cite{PDG2024}
\begin{equation}
\alpha= \arg\!\left[\text{-}\frac{V_{td}V_{tb}^{*}}{V_{ud}V_{ub}^{*}}\right]\! , \quad \quad
  \beta= \arg\!\left[\text{-}\frac{V_{cd}V_{cb}^{*}}{V_{td}V_{tb}^{*}}\right] \!,\quad \quad
  \gamma= \arg\!\left[\text{-}\frac{V_{ud}V_{ub}^{*}}{V_{cd}V_{cb}^{*}}\right]\!,
  \label{eq:angles}
\end{equation}
satisfying $\alpha+\beta+\gamma = \pi$.

\subsection{Experimental determination of $\alpha$, $\beta$ and $\gamma$} 

The angle $\beta$ is the most precisely determined angle of the unitarity triangle, through time-dependent CP asymmetries in $b\to c\bar{c}s$ transitions~\footnote{There is a four-fold ambiguity in the extraction of $\beta$ from determinations of $\sin(2\beta)$ that can be resolved by the measurement of $\cos(2\beta)$ and by a global fit~\cite{Charles:2004jd,UTfit:2005ras}.}. The angle $\gamma$ is instead extracted from time-dependent CP asymmetries in $B\to DK$ processes. Statistical combinations of several decay channels lead to the results~\cite{ParticleDataGroup:2026aaa} 
\begin{equation}
  \beta = \bigl(22.6 \pm 0.5 \bigr)^{\circ}\,,\quad \gamma = \bigl(66.4^{+2.7}_{-2.8} \bigr)^{\circ}\,.
\end{equation}

The angle $\alpha$ can be directly accessed only through $b\to u\bar{u}d$ transitions. The determination is complicated by penguin amplitudes, which have a similar magnitude but a different phase than the tree-level contributions. To date, we have four independent determinations, from time-dependent CP asymmetries in $B^{0}\to\pi\pi$, $B^{0}\to\rho\rho$, $B^{0}\to\rho\pi$ and $B^{0}\to a_1\pi$ decays. The world averages per decay channel obtained by Ref.~\cite{HFLAV:2022esi} are
\begin{equation}
\begin{aligned}
    B^{0}\to\pi\pi\,:&\quad (84.8^{+22.1}_{-5.6})^\circ \cup (99.6^{+7.3}_{-20.4})^\circ\,,\\
    B^{0}\to\rho\rho\,:&\quad (91.0\pm 5.5)^\circ\,,\\
\end{aligned}
\quad \quad \quad  \quad 
\begin{aligned}
    B^{0}\to\rho\pi\,:&\quad (53.4^{+8.3}_{-11.1})^\circ\,,\\
    B^{0}\to a_1\pi\,:&\quad (79\pm 7 \pm 11)^\circ\,.
\end{aligned}
\end{equation}
\begin{table}[t]
\centering
\begin{tabular}{llccc}
\hline
\multicolumn{2}{l}{Experiment} & $\alpha$ & $\beta$ & $\gamma$ \\
\hline
\\[-0.1cm]
\multirow{2}{*}{Belle II~\cite{Belle-II:2018jsg} }&
 $10\,\mathrm{ab}^{-1}$  
  & $2.5^\circ$
  & $0.4^\circ$ 
  & $2.2^\circ$ \\
& $50\,\mathrm{ab}^{-1}$ 
  & $0.6^\circ$ 
  & $0.3^\circ$ 
  & $1.0^\circ$ \\[0.2cm]
\multirow{2}{*}{LHCb~\cite{LHCb:2018roe, ATLAS:2025lrr, LHCb:2021glh, Vagnoni:2025qfv} }
 & $50\,\mathrm{fb}^{-1}$ 
  & - 
  & $0.2^\circ$
  & $0.8^\circ$ \\
&$300\,\mathrm{fb}^{-1}$ 
  & - 
  & $0.08^\circ$ 
  & $0.3^\circ$ \\

\hline
\end{tabular}
\caption{Expected sensitivities for the CKM angles, with the data samples collected in the early 2030s and 2040s by the Belle II and the LHCb experiments.}
\label{tab:future_exp}
\end{table}
Combining these measurements, Ref.~\cite{ParticleDataGroup:2026aaa} obtains a world average of 
\begin{equation}
  \alpha = \bigl(84.1\,{}^{+3.7}_{-3.0}\bigr)^{\circ}\,.
  \label{eq:alpha_exp}
\end{equation}
For more details and results obtained with different statistical treatments we refer the reader to Refs.~\cite{HFLAV:2022esi,Charles:2017evz}.\\
An alternative way to determine $\alpha$, is through the unitarity condition in Eq.~\ref{eq:UT_relation}, from the measurement of the angles $\beta$ and $\gamma$. Using these result and the unitarity of the CKM matrix, we obtain
\begin{equation}
  \alpha^{\rm UT} = \bigl(91.0\,{}^{+2.8}_{-2.8}\bigr)^{\circ}\,.
\end{equation}
Finally, a third determination of $\alpha$ can be obtained by a global fit of the CKM matrix. To the best of our knowledge, some of the most recent results are:
\begin{equation}
  \text{ CKMfitter~\cite{CKMfitter2025}}:\quad \alpha = \bigl(90.82\,{}^{+1.56}_{-0.82}\bigr)^{\circ}\,,\quad \quad \quad 
  \text{ UTfit~\cite{UTfit2025}}:\quad \alpha = \bigl(91.7\,\pm {+1.3}\bigr)^{\circ}\,.
  \label{eq:alpha_fit}
\end{equation}
Although the direct measurements of $\alpha$, $\beta$, and $\gamma$ are not perfectly compatible with CKM unitarity, our fits impose exact unitarity of the CKM matrix. Consequently, the fitted angles represent the values that best accommodate the direct measurements under the unitarity constraint.  
In the future, the Belle II and the LHCb experiment are expected to notably improve on the measurements of $\alpha$, $\beta$ and $\gamma$. Tab.~\ref{tab:future_exp} shows projected sensitivities with data samples that will be recorded by early 2030s and 2040s.

\subsection{Renormalization Group Running of the CKM Matrix}
\label{app:CKM_RGE}
 
In the main text, we argue on the significance of the close proximity of the angles of the unitarity triangle to simple fractions of $\pi$ as a hint for some underlying structure in the Yukawa matrices connected to some UV origin of flavor. However, if there is such a UV explanation it would only determine the angles at the UV scale, $\Lambda_f$, where flavor is generated. It is therefore important to understand the scale dependence of $(\alpha,\beta,\gamma)$. If the RG scaling in the angles was significant, this would force $\Lambda_f$ to be very close to the weak-scale. But very nicely, as we now show, the opposite is true -- $(\alpha,\beta,\gamma)$ barely run at all, and thus the angles we measure in the IR are reflecting what is determined in the UV to extremely high accuracy, well beneath current and future error bars. \\

We work in the $\overline{\rm MS}$ scheme. Let us define 
\begin{equation}
\xi_u = Y_u Y_u^\dagger, \quad     \xi_d = Y_d Y_d^\dagger, \quad \text{ with eigenstates } \quad \xi_u \ket{u_i} = y_{u_i}^2 \ket{u_i}, \, \xi_d \ket{d_i} = y_{d_i}^2 \ket{d_i} \, ,
\end{equation}
so that $V_L^{u,d}$ are the diagonalizing matrices of $\xi_{u,d}$; and the CKM matrix, $V$ in \eqref{eq:CKM_def}, can be equivalently given as $V \equiv \bra{u_i} \ket{d_j}$. To obtain the RGE of the unitarity angles we start by deriving the RGE for the full CKM: 
\begin{equation}
   \dot{V}_{u_i,d_j}
    =
    \langle \dot{u}_i \vert d_j \rangle
    +
    \langle u_i \vert \dot{d}_j \rangle \, ,
    \label{eq:VCKMdot}
\end{equation}
where a dot denotes $d/d\ln\mu$. From first order perturbation theory in quantum mechanics, we know that for any hermitian operator, $\mathcal{O}(\mu)$, with eigenvectors, $\vert \dot{\psi}(\mu) \rangle$ and eigenvalues $\lambda(\mu)$; we can find the $\mu$-dependence of $\vert \dot{\psi}(\mu) \rangle$ in terms of the $\mu$-dependence of $\mathcal{O}$ as follows: 
\begin{equation}
    \vert \dot{\psi}(\mu) \rangle = \sum_{\psi^\prime \neq \psi} \frac{\vert \psi^\prime (\mu)\rangle \langle \psi^\prime(\mu) \vert \dot{\mathcal{O}}(\mu) \vert \psi(\mu) \rangle }{(\lambda(\mu)-\lambda^\prime(\mu))} + c(\mu) \vert \psi(\mu)\rangle, 
\end{equation}
where since $\bra{\psi}\ket{\psi}=1 \Leftrightarrow \langle \dot{\psi} \vert \psi \rangle + \langle \psi \vert \dot{\psi} \rangle =0$, and therefore we have that $c(\mu)$ is purely imaginary, and so we can always rephase $\psi$ to remove $c(\mu)$. As a result, in our case this yields 
\begin{equation}
    \vert \dot{u}_i \rangle =  \sum_{k \neq i} \frac{\vert u_k \rangle \langle u_k \vert \dot{\xi}_u \vert u_i \rangle }{ y_{u_i}^2 - y^2_{u_{k}}}\, , \quad \quad  \vert \dot{d}_j \rangle =  \sum_{k \neq j} \frac{\vert d_k \rangle \langle d_k \vert \dot{\xi}_d \vert d_j \rangle }{ y_{d_j}^2 - y^2_{d_{k}}}. 
    \label{eq:timedep}
\end{equation}
\begin{figure}[t]
    \centering
    \includegraphics[width=\linewidth]{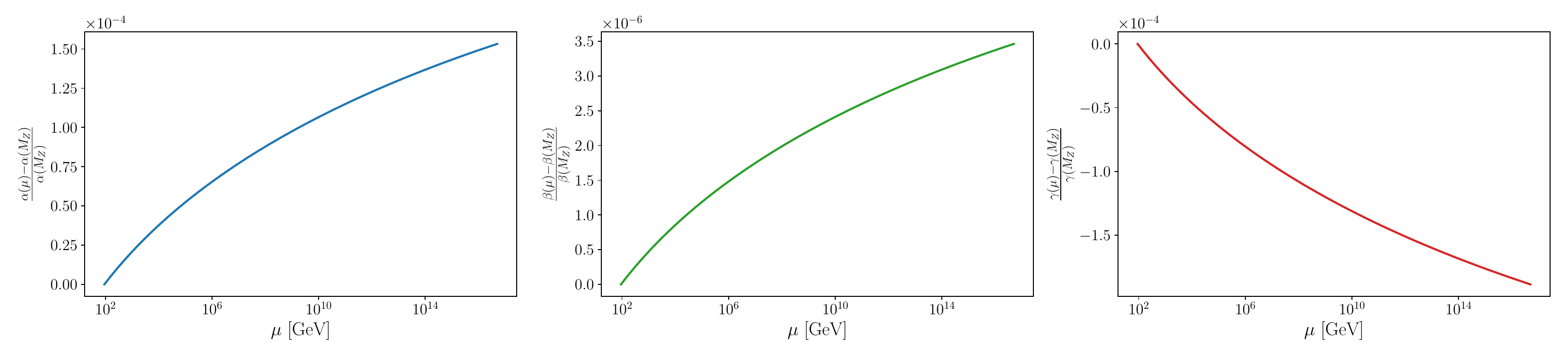}
    \caption{RG evolution of the unitarity angles $(\alpha,\beta,\gamma)$ at 1-loop in the SM.}
    \label{fig:alpha_RGE}
\end{figure}

From the one-loop running of the Yukawa matrices in the SM, we have
\begin{equation}
    16 \pi^2 \dot{Y}_{u} = \alpha_{u} Y_{u} + \beta_u  Y_{u} Y_{u}^\dagger Y_{u} + \gamma_u Y_{d} Y_{d}^\dagger Y_{u}, \quad \Rightarrow \quad   16\pi^2 \dot{\xi}_u = (\alpha_{u}+\alpha^*_{u}) \xi_{u} + (\beta_u+\beta^*_u ) \xi_u^2 + (\gamma_u \xi_d \xi_u + \text{h.c.}),
\end{equation}
and similar for the down sector. So the only 1-loop contribution which leads to a non-vanishing contribution to \eqref{eq:timedep} is then the $\gamma_{u,d}$ piece. Plugging it back into \eqref{eq:VCKMdot}, we find the following 1-loop RG equation for the CKM matrix:
\begin{equation}
    16 \pi^2 \dot{V}_{i,j} = \gamma_{u}  \sum_{l} \sum_{k\neq i }  r^u_{i,k} \lambda_{d_l}^2 V_{i,l} V^*_{k,l} V_{k,j} + \gamma_d \sum_{l} \sum_{k\neq j }  r^d_{j,k} \lambda_{d_l}^2 V_{i,k} V^*_{l,k} V_{l,j}, \quad \text{ with } r_{i,k} = \frac{y_{i}^2 + y_{k}^2}{y_{i}^2 - y_{ k}^2} \, ,
\end{equation}
where $\gamma_{u,d} = \gamma^*_{u,d} = -3/2$. From this, together with the definition of $(\alpha,\beta,\gamma)$ in terms of the $V$, as given in \eqref{eq:angles}, we can derive the 1-loop RG evolution for the angles -- the results are shown in Fig. \ref{fig:alpha_RGE}. As we can see, even up to the GUT scale, the angles change at most by $\sim 0.01\%$! This is striking simply because the individual entries of $V$ run a non-negligible amount, but still the ratios defining the unitarity triangle remain almost invariant.

We can understand this result in a simple way analytically.  Keeping the leading contribution to the RGE's, it is very easy to see that, up to miniscule corrections, the sides of the unitarity triangle are rescaled uniformly under the RG, and thus the shape of the triangle (and hence the angles) does not change. The largest corrections to the RG come from the top Yukawa $y_t$; the other corrections are already much smaller than current or projected future experimental bounds. 
Then, we have that 
\begin{equation}
16\pi^2 \dot{V}_{i, j} = \gamma_d \sum_{k \neq j} r^d_{j,k} y_t^2 \, V_{i,k} V^*_{3,k} V_{3,j} \, . 
\end{equation}
Since the down eigenvalues are hierarchical we can replace $r^d_{j,k} \to +1$ when $j>k$ and $r^d_{j,k} \to -1$ when $j<k$. Hence, looking at the running of $V_{1,3}$ and $V_{3,1}$ that determine the unitarity triangle, we have 
\begin{equation}
16 \pi^2 \dot{V}_{1,3}=\gamma_d \,  y_t^2 \sum_{k \neq 3} V_{1,k} V^*_{3,k} V_{3,3} \times (+1) = - \gamma_d y_t^2 V_{1,3} V^*_{3,3} V_{3,3}\, ,
\end{equation}
where to get the second form we used unitarity of $V$. Similarly we find
\begin{equation}
16\pi^2 \dot{V}_{3,1}=\gamma_d \, y_t^2 \sum_{k \neq 1} V_{3,k} V^*_{3,k} V_{3,1} \times (-1) = -\gamma_d y_t^2 (1 - V_{3,1} V^*_{3,1}) V_{3,1}\, .
\end{equation}
Since $|V_{3,3}|^2$ and $(1 - |V_{3,1}|^2)$ are both incredibly close to 1, we find that to leading order
\begin{equation}
16 \pi^2\dot{V}_{1,3} = - \gamma_d y_t^2 V_{1,3}, \quad \quad \, 16 \pi^2 \dot{V}_{3,1} = -\gamma_d y_t^2 V_{3,1}
\end{equation}
which tells us that at leading order, the sides of the unitarity triangle are rescaled by exactly the same factor under the RG, hence the angles remain fixed. 

\section{Approximate Analytic Predictions for $\alpha, \beta, \gamma$ from Texture Zeros}
\label{app:U(1)9}

We start by considering some underlying theory of flavor where all entries of $Y^{U,D}$ are non-zero and complex. In general $\alpha, \beta, \gamma$ will have a complicated dependence on the 10 phases that cannot be removed by field redefinitions. In view of the experimental observation that $\alpha \simeq \pi/2,$, $\beta \simeq \pi/8,$ and $\gamma \simeq 3\pi/8$, are there any simple constraints on the theory that force $\alpha, \beta, \gamma$ close to multiples of $\pi/8$?  We assume that
the small dimensionless numbers observed in quark mass ratios and the CKM matrix are accounted for by $Y_{u,d}$ taking a hierarchical form
\begin{equation}
  |Y^{u,d}_{ij}| \ll |Y^{u,d}_{kl}|, \qquad i+j < k + l.
  \label{eq:Yhier}
\end{equation}
This still leaves great ambiguity in the magnitudes of the various elements and, on occasions, it is useful to have a more definite standard hierarchical form.  We use a form governed by an approximate $U(1)^9$ flavor symmetry \cite{Carrasco-Martinez:2026wzu}, with one U(1) and one real spurion, $\epsilon^{q, u^c,d^c}_i$, for each of the 9 Weyl quark fields, giving 
\begin{equation}
       Y^u_{ij} \; = \; C^u_{ij} \, \epsilon^q_i \epsilon^{ u^c}_j,\hspace{0.2in}
       Y^d_{ij} \;=\; C^d_{ij} \, \epsilon^q_i \epsilon^{d^c}_j.
\label{eq:U(1)9}
\end{equation}
Fitting to the quark masses and mixing angles, $\epsilon^{q, u^c, d^c}_i$ are chosen so that $C^{u,d}_{ij}$ have order unity magnitudes and arbitrary phases. A texture zero occurs in $Y^{u,d}_{ij}$ if $|C^{u,d}_{ij}| \ll 1$.\\ 
In a 2-generation model, $Y_{u,d}$ can be diagonalized exactly by small-angle bi-unitary transformations, giving the CKM matrix $V= V^u_LV^{d\,\dagger}_L$ with
\begin{equation}
V^{u,d}_L =  \begin{pmatrix}
c & s  \\
-s^* & c 
\end{pmatrix}^{u,d} = \left[P^* \begin{pmatrix}
c & |s|  \\
-|s| & c 
\end{pmatrix} P\right]^{u,d}; \qquad s^{u,d} = |s|^{u,d}\;e^{i\phi^{u,d}}, \; \;c^{u,d} = \sqrt{1 - (|s|^2)^{u,d}}, 
\label{eq:V2X2}
\end{equation}
where $P^{u,d}$ is a diagonal phase matrix Diag$(e^{i \phi^{u,d}}, 1)$ and
\begin{equation}
  |s|^{u,d} \simeq  -\left|\frac{Y^{u,d}_{12}}{Y^{u,d}_{22}}\right|,  \qquad
  \phi^{u,d} = \text{arg}\left[\frac{Y^{u,d}_{12}}{Y^{u,d}_{22}} \left(1+ \frac{Y^{*u,d}_{21} Y^{u,d}_{11}}{Y^{u,d}_{12} Y^{*u,d}_{22}} \right) \right] \simeq \text{arg}\left(\frac{Y^{u,d}_{12}}{Y^{u,d}_{22}} \right).
  \label{eq:phi2X2}
\end{equation}

With three generations, $Y_{u,d}$ can be approximately diagonalized by successive small-angle $2 \times 2$ bi-unitary transformations; the CKM matrix is approximately 
\begin{equation}
  V \simeq V^u_{12} \; V^u_{13}\; V^u_{23} \;V^{d\dagger}_{23}\;V^{d\dagger}_{13}\;V^{d\dagger}_{12}
  \label{eq:V3X3}
\end{equation}
with 
\begin{align}
s^{u,d}_{23}  \simeq  \frac{Y^{u,d}_{23}}{Y^{u,d}_{33}}, \;\; s^{u,d}_{13}  \simeq  \frac{Y^{u,d}_{13}}{Y^{u,d}_{33}}; \qquad \;s^{u,d}_{12}  \simeq  \frac{Y'^{u,d}_{12}}{Y'^{u,d}_{22}}, \;\;\; 
Y'^{u,d}_{12}  \simeq Y^{u,d}_{12} - \frac{Y^{u,d}_{32}}{Y^{u,d}_{33}}Y^{u,d}_{13}, \;\;\;
  Y'^{u,d}_{22}  \simeq Y^{u,d}_{22} - \frac{Y^{u,d}_{32}Y^{u,d}_{23}}{Y^{u,d}_{33}}
  \label{eq:sij}
\end{align}
at leading order, giving
\begin{equation}
  V \simeq  \begin{pmatrix}
1 & s_{12} \; &  s_{13} + s_{12}^{u} s_{23}\\
-s_{12}^* & 1 & s_{23} \\ 
-s_{13}^* - s_{12}^{d*} s_{23}^*\; & -s_{23}^* & 1
\end{pmatrix}, \quad \quad  s_{ij} = s^u_{ij} - s^d_{ij}.
\label{eq:Vgen}
\end{equation}
Since $|V_{13}| \ll |V_{12}|, |V_{23}|$, we have ignored $s_{13}^{u,d} s_{23}$ terms in $V_{12, 21}$ and $s_{12}^{u,d} s_{13}$ terms in $V_{23, 32}$. However, including these terms does not affect $\alpha$. 
The key leading order predictions for $\alpha, \beta, \gamma$ are
\begin{equation}
  \alpha \; \simeq \; \,\text{arg}\left(\frac{s_{13}^* + s_{12}^{d*} s_{23}^*}{s_{13}^* + s_{12}^{u*} s_{23}^*}\right), \qquad \beta \; \simeq \; \,\text{arg}\left(\frac{s^{d*}_{12} s_{23}^* - s^{u*}_{12} s_{23}^*}{ s_{12}^{d*} s_{23}^* + s_{13}^*}\right), \qquad \gamma \; \simeq \; \,\text{arg}\left(\frac{s_{13}^* + s_{12}^{u*} s_{23}^*}{s^{u*}_{12} s_{23}^* - s^{d*}_{12} s_{23}^*}\right).
  \label{eq:abc}
\end{equation}
Inserting $s_{ij}^{u,d}$ from (\ref{eq:sij}) gives $\alpha, \beta, \gamma$ in terms of the Yukawa matrix elements. \\
In the absence of texture zeros, there is no reason for any angle of the unitarity triangle to be close to a multiple of $\pi/8$.
Using the hierarchical forms of (\ref{eq:U(1)9}), corresponding to approximate $U(1)^9$ flavor symmetry, $s_{13}$, $s_{12}^{d} s_{23}$ and $s_{12}^{u} s_{23}$ have comparable magnitudes of $\epsilon^q_1/\epsilon^q_3$.  Consider a theory where CP and flavor symmetries are spontaneously broken via the vevs of a set of flavon fields having phases that are multiples of $\pi/8$.  Further, assume that each entry in $Y^{u,d}_{ij}$ is dominated by products of these vevs so that, to leading order, each $Y^{u,d}_{ij}$ and each $s^{u,d}_{ij}$ has a phase that is a multiple of $\pi/8$. Could it be that, by choosing all terms in each numerator and denominator of (\ref{eq:abc}) to have the same phase, the ratios then predict angles of the unitarity triangle that are multiples of $\pi/8$?  The answer is {\it no}; the expressions for $\alpha, \beta$ and $\gamma$ each have a common term in the numerator and denominator ($s_{13}, s^d_{12} s_{23}$ and $ s^u_{12} s_{23}$ respectively), so that this procedure forces $\alpha, \beta, \gamma =0$. 

For $\alpha, \beta, \gamma$ to have a direct connection to the phases of the Yukawa matrix elements, the numerators and denominators of (\ref{eq:abc}) should not contain the same term. There are three extremely simple ways this can happen. Although there are six mixings, $s^{u,d}_{ij}$, the angles $\alpha, \beta$ and $\gamma$ depend on only four combinations, $s_{12}^u, s_{12}^d, s_{13}$ and $s_{23}$. While $V_{cb}$ requires $s_{23} \neq 0$, one of $s_{12}^u, s_{12}^d, s_{13}$ can vanish while still agreeing with data. Each of these three cases gives a greatly simplified expression for one of the angles of the unitarity triangle
\begin{equation}
\alpha \simeq \text{arg}\left(\frac{s_{12}^{u}}{s_{12}^{d}}\right), \;\; s_{13}\simeq 0;
\qquad
 \beta \simeq \text{arg}\left(-\frac{s_{13}}{s_{12}^u s_{23}}\right), \;\; s_{12}^d\simeq 0; \qquad
  \gamma \simeq \text{arg}\left(-\frac{s_{12}^d s_{23}}{s_{13}}\right), \;\; s_{12}^u\simeq 0.
  \label{eq:abcsimple}
\end{equation}
Thus, spontaneous CP violation leads to a natural explanation for why one of the unitarity triangle angles is a multiple of $\pi/8$ if texture zeros force one of $s_{12}^u, s_{12}^d, s_{13}$ to vanish. Furthermore, when the expressions in (\ref{eq:abcsimple}) are written in terms of Yukawa matrix elements, using (\ref{eq:sij}), they should either be monomials in $Y^{u,d}_{ij}$ or, if there is a sum of terms, each term should have the same phase. For example, the two texture zeros $|C^{u,d}_{13}| \ll 1$ give
\begin{equation}
Y^{u,d}_{13} \simeq 0, \qquad \alpha \; \simeq \; \,\text{arg}\left(\frac{Y^d_{22} - \frac{Y^d_{23}Y^d_{32}}{Y^d_{33}}}{Y^d_{12}} \; \frac{Y^u_{12}}{Y^u_{22} - \frac{Y^u_{23}Y^u_{32}}{Y^u_{33}}}\right).
  \label{eq:alphaY13=0}
\end{equation}
If there are two further texture zeros from $C^{u,d}_{32} \ll 1$, then $\alpha$ is the argument of a monomial and, if spontaneous CP violation leads to each matrix element being dominantly real or imaginary, there is a 50\% chance of $\alpha \simeq \pi/2$. The same result occurs without further texture zeros if $Y^{u,d}_{22}$ have the same phases as $Y^{u,d}_{23}Y^{u,d}_{32}/Y^{u,d}_{33}$. 
Similarly, there are many possibilities for spontaneous CP violation to yield $\beta \simeq \pi/8$ or $\gamma \simeq 3\pi/8$. In these cases, it may be more likely if the expressions in (\ref{eq:abcsimple}) are monomials rather than sums of terms with the same phases. There are several different textures for each of these cases.

An alternative to (\ref{eq:abcsimple}) arises if there are relations among various $s^{u,d}_{ij}$. Such relations occur when two $Y_{ij}$ have a texture zero but both are generated by the same rotation on right-handed quark fields. For example, even with $Y^d_{12,22} = 0$, $Y^d_{32}$ can generate $s^d_{12} = -s^d_{13}/s^d_{23}$. Substituting into (\ref{eq:abc}) for $\beta$, and setting $s^u_{23}=0$ by a texture zero, the numerators and denominators no longer contain the same term. A similar result occurs for $\gamma$ when the relation is in the up sector, thus
\begin{equation}
  \beta \simeq \text{arg}\left(\frac{s^u_{13}}{s_{12} s^d_{23}}\right), \;\; s^d_{13} \simeq -s^d_{12} s^d_{23}, \; s^u_{23}\simeq 0;  \qquad
  \gamma \simeq \text{arg}\left(\frac{s_{12} s^u_{23}}{s^d_{13}}\right), \;\; s^u_{13} \simeq -s^u_{12} s^u_{23}, \; s^d_{23}\simeq0.
  \label{eq:bccancel}
\end{equation}
Either $s^d_{13} \simeq -s^d_{12} s^d_{23}$ or $s^u_{13} \simeq -s^u_{12} s^u_{23}$ can lead to $\alpha$ not having identical terms in numerator and denominator, but only in the presence of two extra texture zeros
\begin{equation}
\alpha \simeq \text{arg}\left(\frac{s_{23}^d}{s_{23}^u}\right), \qquad s^u_{13} \simeq -s^u_{12} s^u_{23}, \; s^d_{13}\simeq 0, \; s^d_{12}\simeq 0 \qquad  or \qquad s^d_{13} \simeq -s^d_{12} s^d_{23}, \; s^u_{13}\simeq 0, \; s^u_{23}\simeq 0.
  \label{eq:acancel}
\end{equation}
Thus, in general there are many ways that textures and spontaneous CP violation can lead to one angle of the unitarity triangle being a multiple of $\pi/8$. However, such phase arguments cannot explain why a second angle also has this form. 
We conclude that (\ref{eq:npi/8}) {\it hints that CP is spontaneously broken in nature and that the Yukawa matrices have texture zeros}, entries that are highly suppressed relative to expectations from approximate Abelian flavor symmetries. Future data could greatly strengthen this conclusion.

Even when texture zeros are present, there are several origins for deviations of $\alpha, \beta$ or $\gamma$ from a multiple of $\pi/8$. First, diagonalization by successive $2 \times 2$ transformations is approximate, so that there are corrections to (\ref{eq:V3X3}) and (\ref{eq:sij}). Most important for theories with $Y^{u,d}_{13} = 0$, the transformations $Y^{u,d}_{23}$ of (\ref{eq:V3X3}) regenerate small but non-zero values of $Y^{u,d}_{13}$. Also, the exact results for the phases $\phi_{ij}$, which take forms analogous to (\ref{eq:phi2X2}), deviate from the leading result of 
$\text{arg}(Y_{ij}/Y_{jj})$. However, with $U(1)^9$ approximate flavor symmetries these deviations are suppressed by $m_i/m_j$, and are small. In general, none of these deviations can be predicted; even using $U(1)^9$ as a guide, there are too many unknown parameters in $C^{u,d}_{ij}$. Hence, we now discuss the highly predictive set of theories with 9 texture zeros, the main subject of this paper.

Textures with 9 links have 9 real parameters and a single phase, $\varphi$. Six of the 9 real parameters are determined by the quark mass eigenvalues, leaving three to describe the six mixings $|s^{u,d}_{ij}|$. In all 9-link textures some of these mixings are zero because of the texture zeros. Indeed, most have three of them zero, with the three that are non-zero chosen to fit the data. Requiring $V_{cb} \neq 0$ requires $s^u_{23}$ or $s^d_{23}$ to be non-zero. There are three realistic possibilities for the other two non-zero mixings, one has both $s^{u,d}_{12}$ non-zero and $s_{13} =0$, while the others have one of $s^{u,d}_{12}$ non-zero and one of $s^{u,d}_{13}$ non-zero. Each possibility leads to a simple prediction for one of the angles of the unitarity triangle
\begin{equation}
\alpha \simeq \text{arg}\left(\frac{s_{12}^u}{s_{12}^d}\right), \;\; s_{13}\simeq 0;
\qquad
 \beta \simeq \text{arg}\left(-\frac{s_{13}}{s_{12}^u s_{23}}\right), \;\; s_{12}^d\simeq 0; \qquad
  \gamma \simeq \text{arg}\left(-\frac{s_{12}^d s_{23}}{s_{13}}\right), \;\; s_{12}^u\simeq 0,
  \label{eq:abc9}
\end{equation}
where, in the expressions for $\beta$ and $\gamma$, it is understood that only one of $s^{u,d}_{13}$ and one of $s^{u,d}_{13}$ are non-zero. These results have a close similarity with the more general result of (\ref{eq:abcsimple}). A key difference is that, for the 9-link textures, one angle of the unitarity triangle is given by the phase $\varphi$, as the expressions involve the unique rephasing-invariant.  All these textures lie in the $\varphi = \pi/2, \pi/8$ and $3 \pi/8$ peaks of Fig. \ref{fig:theta_histogram}.

There are also 9-link textures where 4 $s^{u,d}_{ij}$ are non-zero. Since they are described by three free parameters only three are independent, so there is a relation between them. If $Y_{12,22} = 0$, a rotation on right-handed quarks to remove $Y_{32}$ can simultaneously generate $Y_{12,22}$ from $Y_{13,23}$, giving
\begin{equation}
s^d_{12} \; = \; -\frac{s^d_{13}}{s^d_{23}} \qquad {\rm or} \qquad s^u_{12} \; = \; -\frac{s^u_{13}}{s^u_{23}}.
  \label{eq:srel1}
\end{equation}
which lead to the results (\ref{eq:bccancel}) and (\ref{eq:acancel}), with $s^u_{12}\simeq 0$ and $s^d_{12} \simeq 0$ in the results for $\beta$ and $\gamma$, respectively. Again, one angle of the unitarity triangle is given by the phase $\varphi$,  so these textures also populate the peaks of Fig. \ref{fig:theta_histogram}. If $Y^d_{12,23} = 0$, a rotation on right-handed quarks to remove $Y^d_{32}$ can simultaneously generate $Y^d_{12,23}$ from $Y^d_{13,22}$, giving
\begin{equation}
s_{12}^d \simeq -s_{23}^d s^d_{13} \left(\frac{y_b}{y_s} \right)^2, \qquad s^u_{23} \neq 0.
  \label{eq:pi/4rel}
\end{equation}
These textures populate the peak near $\pi/4$ in Fig.~ \ref{fig:theta_histogram}. It is an accident that $\varphi$ is near a multiple of $\pi/8$; as $y_b/y_s$ deviates from the observed value, the peak in $\varphi$ moves away from $\pi/4$, as we show in App.~\ref{app:fit-free}. A similar relation with up and down sectors interchanged is not realistic. 

For hierarchical Yukawa matrices with 9-link textures, the only realistic relations between $s^{u,d}_{ij}$ are given by (\ref{eq:srel1}) and (\ref{eq:pi/4rel}). For (\ref{eq:pi/4rel}) there is no connection between $\varphi$ and $(\alpha, \beta, \gamma)$, which are accidentally close to multiples of $\pi/8$. However, {\it all other textures lead to  (\ref{eq:abc9}) or (\ref{eq:srel1}) at leading order requiring one of $(\alpha, \beta, \gamma)$ to be equal to the rephasing-invariant angle $\varphi$.} This explains both the peaks in Fig. \ref{fig:theta_histogram} and the absence of textures between the peaks. 

\subsection{Beyond 9-link textures: numerical sensitivity}
\begin{figure}[t]
    \centering
\includegraphics[width=0.6\linewidth]{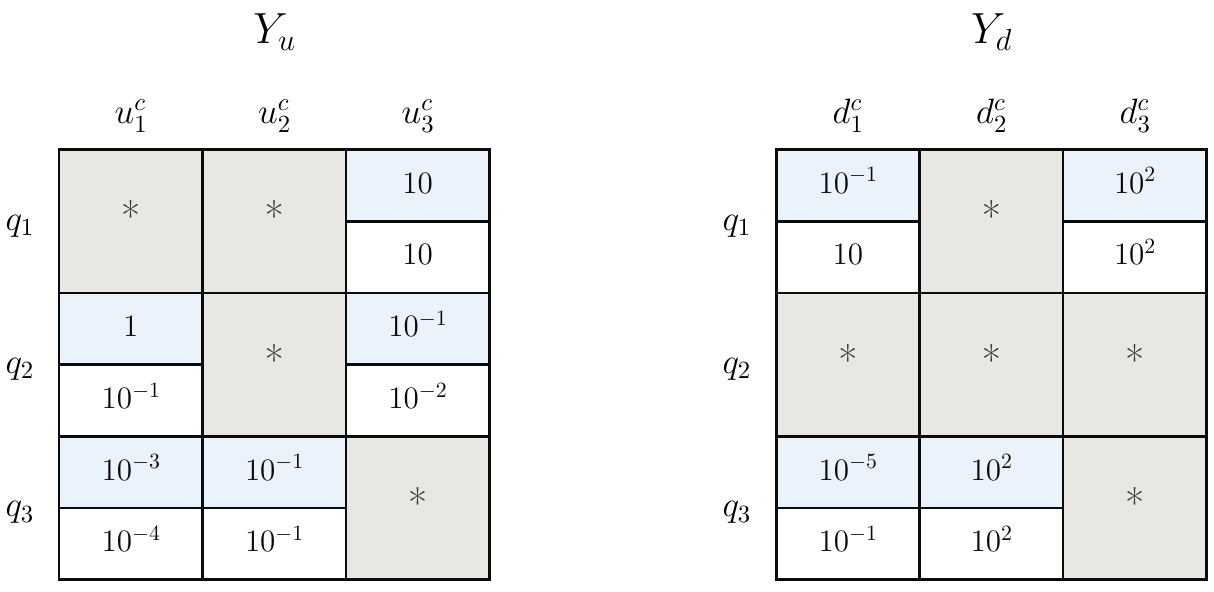}
    \caption{Sensitivity of the $\alpha\simeq\pi/2$ prediction to each of the nine vanishing entries of the texture in Fig.~\ref{fig:IntroFigure} of the main text. Gray cells (marked $*$) are the non-zero entries fixed by the fit to quark masses and CKM observables. For each vanishing entry, the order of magnitude, in degrees, of the derivative in Eq.~\eqref{eq:sensitivity} is shown, evaluated for a purely imaginary (top, light blue) and purely real (bottom, white) perturbation, retaining the smaller of the two signs in each case.}
    \label{fig:SensitivityPlot}
\end{figure}

Having established that, at leading order, all 9-link textures, apart from those with $\varphi$ near $\pi/4$, require one of the unitarity-triangle angles $(\alpha,\beta,\gamma)$ to coincide with the rephasing-invariant phase $\varphi$, we now ask how robust this prediction is under the addition of further non-vanishing Yukawa entries. More precisely, we would like to understand how many zeros can be removed while preserving the predictions $\alpha\simeq\pi/2$, $\beta\simeq\pi/8$, or $\gamma\simeq3\pi/8$ as genuine consequences of the texture, rather than accidental numerical coincidences. We illustrate the general idea by analyzing the texture shown in Fig.~\ref{fig:IntroFigure} of the main text. We begin with the corresponding 9-link texture, fixing its non-vanishing entries to the values obtained from the fit to quark masses and CKM observables, and the rephasing invariant phase to $\pi/2$. This texture predicts $\alpha\simeq\pi/2$ at leading order. We then switch on the vanishing entries, one by one, at their expected hierarchical size, $Y^{u/d}_{ij}\sim\sqrt{\epsilon_i^q\,\epsilon_j^{u^c/d^c}}$, where the $\epsilon$'s denote the $U(1)^9$ spurions, and compute the numerical derivative of $\alpha$,
\begin{equation}
\lim_{\varepsilon\to0}
\frac{\alpha\!\left[Y^{u/d}_{ij}
=\varepsilon\sqrt{\epsilon_i^q\,\epsilon_j^{u^c/ d^c}}\right]
-\alpha\!\left[Y^{u/d}_{ij}=0\right]}
{\varepsilon}.
\label{eq:sensitivity}
\end{equation}
Turning on a previously vanishing entry introduces an additional rephasing invariant into the theory. Consequently, the shift in $\alpha$ depends not only on the magnitude of the new Yukawa entry but also on its phase. To account for this dependence, we evaluate the derivative above for $Y_{ij}$ taken to be purely real or purely imaginary, with both positive and negative signs, and retain the smallest value in each case. 

The results are presented in Fig.~\ref{fig:SensitivityPlot}. Each zero entry of the texture is assigned two numbers: the smallest derivative obtained for a purely imaginary perturbation (top) and for a purely real perturbation (bottom), both expressed in degrees. It is immediately apparent that three entries, $Y^u_{13}$, $Y^d_{13}$, and $Y^d_{32}$, must be suppressed by one to two orders of magnitude with respect to their natural values, independently of their phase, in order to preserve the prediction $\alpha\simeq\pi/2$ at the level of one degree. The same conclusion applies to $Y^d_{11}$ when the perturbation is purely real, and to $Y^u_{21}$ when it is purely imaginary, once the projected LHCb precision with $300\,\mathrm{fb}^{-1}$ for $\beta$ and $\gamma$ and the assumption of CKM unitarity is taken into account (see Tab.~\ref{tab:future_exp}). 

We checked that similar conclusions hold for several other textures, and in particular all the nine classes with $\varphi=\pi/2$ in Fig.~\ref{fig:PhaseClasses}. This is consistent with the expectation that, as the number of texture zeros decreases, the unitarity triangle angles receive multiple leading-order contributions, generally carrying different phases.
Overall, this analysis indicates that at least four texture zeros are required to ensure that the relation between the unitarity-triangle angles and the rephasing-invariant phase is a genuine prediction of the texture, rather than an accidental numerical coincidence.

\section{Perturbative diagonalization}
\label{app:Examples}

In this appendix, we explicitly examine three representative theories, where at leading order $\alpha$, $\beta$ and $\gamma$, respectively, are the argument of a single rephasing invariant monomial. We perturbatively diagonalize the Yukawa matrices, parameterizing each entry as a power of a small parameter $\epsilon$ (informed by the numerical fit), and compute the leading and subleading corrections to the unitarity angles. Since the parameters of the theory are determined by a fit to the physical observables, these corrections are necessarily expressible in terms of CKM matrix elements and quark masses.

\subsubsection*{Example I}

First of all, we analyze a texture with $\varphi=\pi/2$. The Yukawa matrices are
\begin{align}
\begin{split}
    Y_u = \begin{pmatrix}
    0 & i\,u_{12}\,\epsilon^4 & 0\\
    u_{21}\,\epsilon^5 & u_{22}\,\epsilon^3 & 0\\
    0 & 0 & u_{33}\\
    \end{pmatrix}\,,\quad
    Y_d = \begin{pmatrix}
    0 & d_{12}\,\epsilon^5 & 0\\
    d_{21}\,\epsilon^5 & d_{22}\,\epsilon^4 & d_{23}\,\epsilon^3\\
    0 & 0 & d_{33}\,\epsilon^2\\
    \end{pmatrix}
\end{split}
\end{align}
where $\epsilon\sim 0.1$ so that all the $u_{ij}$ and $d_{ij}$ are $\mathcal{O}(1)$ numbers. It is instructive to inspect the matrices $V^u_L$ and $V^d_L$, which diagonalize $Y_uY_u^\dagger$ and $Y_dY_d^\dagger$. $V^u_L$ is a 1-2 rotation with $s_{12}^u\simeq -i\,u_{12}/u_{22}\,\epsilon$. For $V^d_L$ one finds
\begin{align}
\begin{split}
    V^d_L \simeq \begin{pmatrix}
    1 & s_{12}^d & s_{12}^d\, s_{23}^d\\
    -s_{12}^d & 1 & s_{23}^d\\
    \mathcal{O}(\epsilon^6) & -s_{23}^d & 1\\
    \end{pmatrix}\,,
\end{split}
\end{align}
where $s_{12}^d\simeq -d_{12}/d_{22}\,\epsilon$ and $s_{23}^d\simeq -d_{23}/d_{33}\,\epsilon$.
Note the suppression in $(V^d_L)_{31}$ which can be easily understood by looking at the shortest path that connect $Q_3$ and $Q_1$ in $Y_d$: $(V^d_L)_{31} \simeq \epsilon^6 (d_{12}d_{22}d_{23}/d_{33}^3)$. It is now trivial to obtain the leading order contributions to the CKM matrix
\begin{align}
\begin{split}
    V_{\rm CKM} = V_L^{u} V_L^{d\, \dagger} \simeq \begin{pmatrix}
    1 & -s_{12}^d + s_{12}^u & -s_{12}^u\,s_{23}^d\\ s_{12}^d - s_{12}^{u\,*} & 1 & -s_{23}^d\\
    s_{12}^d\, s_{23}^d & s_{23}^d & 1\\
    \end{pmatrix}\,.
\end{split}
\end{align}
Therefore, at leading order $V_{ud}, V_{td}$ and $V_{tb}$ are real, while $V_{ub}$ is purely imaginary and $\alpha = {\rm arg}(-V_{td}V_{tb}^*/V_{ud}V_{ub}^*) = 90^\circ$. The CKM elements and quark yukawas at leading order in $\epsilon$ are 
\begin{align}
\begin{split}
    V_{ud} &\simeq V_{cs} \simeq V_{tb} \simeq 1 + \mathcal{O}(\epsilon^2)\,,\; V_{us} = \epsilon\bigg(\frac{d_{12}}{d_{22}} - i\,\frac{u_{12}}{u_{22}} \bigg) + \mathcal{O}(\epsilon^3)\,,\; V_{ub} = -i\,\epsilon^2\,\frac{d_{23}u_{12}}{d_{33}u_{22}} + \mathcal{O}(\epsilon^4)\,,\\
    V_{cd} &= \epsilon\bigg(-\frac{d_{12}}{d_{22}} - i\,\frac{u_{12}}{u_{22}} \bigg) + \mathcal{O}(\epsilon^3)\,,\; V_{cb} = \epsilon\frac{d_{23}}{d_{33}} + \mathcal{O}(\epsilon^3)\,,\; V_{td} = \epsilon^2\,\frac{d_{12}d_{23}}{d_{33}d_{22}} + \mathcal{O}(\epsilon^4)\,,\; V_{ts} = -\epsilon\frac{d_{23}}{d_{33}} + \mathcal{O}(\epsilon^3)\,,\\
    y_u^2 &\simeq \epsilon^{12} \frac{u_{12}^2u_{21}^2}{u_{22}^2}\,,\;  y_c^2 \simeq \epsilon^{6} u_{22}^2\,,\; y_t^2 \simeq u_{33}^2\,,\; y_d^2 \simeq \epsilon^{12} \frac{d_{12}^2d_{21}^2}{d_{22}^2}\,,\; y_s^2 \simeq \epsilon^8 d_{22}^2\,,\; y_b^2 \simeq \epsilon^{4} d_{33}^2\,.
\end{split}
\end{align}
To compute the next-to-leading order correction to $\alpha$, we need higher order corrections to the CKM elements. We find

\begin{align}
    \begin{split}
       V_{ud} &= 1-\epsilon^2\bigg(\frac{d_{12}^2}{2d_{22}^2} + \frac{u_{12}^2}{2u_{22}^2} - \frac{i\,u_{12}d_{12}}{d_{22}u_{22}}  \bigg) + \mathcal{O}(\epsilon^3)\,,\; V_{ub} = -i\,\epsilon^2\,\frac{d_{32}u_{12}}{d_{33}u_{22}} +i\,\epsilon^4\,\frac{d_{23}u_{12}(d_{33}^2u_{12}^2+d_{23}^2u_{22}^2)}{2d_{33}^3u_{22}^3} + \mathcal{O}(\epsilon^5)\,,\\
       V_{td} &= \epsilon^2\,\frac{d_{12}d_{23}}{d_{33}d_{22}} -\epsilon^4\,\frac{d_{23}d_{12}(d_{12}^2+2d_{21}^2)}{2d_{22}^3d_{33}} + \mathcal{O}(\epsilon^3)\,,\; V_{tb} = 1-\epsilon^2\frac{d_{23}^2}{2d_{33}^2} + \mathcal{O}(\epsilon^5)\,,\\
       \alpha &= \arg\bigg(-\frac{V_{td}V_{tb}^*}{V_{ud}V_{ub}^*}\bigg) \simeq \arg\bigg[i\,\frac{d_{12}u_{22}}{d_{22}u_{12}} + \epsilon^2 \bigg(\frac{d_{12}^2}{d_{22}^2} + i\,\frac{d_{12}d_{22}^2u_{12}^2-d_{12}d_{21}^2u_{22}^2}{d_{22}^3u_{12}u_{22}} \bigg) + \mathcal{O}(\epsilon^3)\bigg] \simeq \frac{\pi}{2} - \frac{d_{12}u_{12}}{d_{22}u_{22}}\epsilon^2 + \mathcal{O}(\epsilon^3)\\
       &\simeq \frac{\pi}{2} - \bigg|\frac{V_{td}\,V_{ub}}{V_{cb}\,V_{ts}}\bigg|\,.
    \end{split}
\end{align}

\subsubsection*{Example II}

Now, we study a texture with $\varphi = -\pi/8$. The Yukawa matrices are
\begin{align}
\begin{split}
    Y_u = \begin{pmatrix}
    u_{11}\,\epsilon^5 & 0 & u_{13}\,\epsilon^2\\
    0 & u_{22}\,\epsilon^2 & 0\\
    0 & 0 & u_{33}\\
    \end{pmatrix}\,,\quad
    Y_d = \begin{pmatrix}
    d_{11}\,\epsilon^5 & 0 & d_{13}\,\epsilon^4\\
    0 & 0 & d_{23}\,\epsilon^3\\
    0 & d_{32}\epsilon^2 & e^{i\varphi}\,d_{33}\,\epsilon^2\\
    \end{pmatrix}\,,
\end{split}
\end{align}
$V_L^d$ is well-approximated by $V_L^d \simeq R_{12}\cdot R_{13}\cdot R_{23}$, with $s^d_{23}\simeq\, -e^{-i\varphi} d_{23}d_{33}/(d_{32}^2+d_{33}^2)\,\epsilon$, $s^d_{13}\simeq\, -e^{-i\varphi}d_{13}d_{33}/(d_{32}^2+d_{33}^2)\,\epsilon^2$ and $s^d_{12}=-d_{13}/d_{23}\,\epsilon$,
\begin{align}
    V^d_L \simeq \begin{pmatrix}
    1 & s^d_{12} & s^d_{13} + s^d_{12}s^d_{23}\\[2mm]
    -s^d_{12} & 1 & s^d_{23}\\[2mm]
    -s^{d\, *}_{13} & -s^{d\, *}_{23} & 1
    \end{pmatrix}\,.
\end{align}
$V_L^u$ is an exact $1$-$3$ rotation, with $s^u_{13} \simeq -u_{13}/u_{33}\,\epsilon^2$.
The CKM elements and quark Yukawas at leading order are
\begin{align}
\begin{split}
    V_{ud} &\simeq V_{cs} \simeq V_{tb} \simeq 1+\mathcal{O}(\epsilon^2)\,,\; V_{us} = \frac{d_{13}}{d_{23}}\,\epsilon + \mathcal{O}(\epsilon^3)\,,\; V_{ub} = \bigg[\,\frac{d_{13}d_{33}}{d_{32}^2 + d_{33}^2}e^{i\varphi}-\frac{u_{13}}{u_{33}}\,\bigg]\epsilon^2 + \mathcal{O}(\epsilon^4)\,,\\[2mm]
    V_{cd} &= -\frac{d_{13}}{d_{33}}\,\epsilon + \mathcal{O}(\epsilon^3)\,,\; V_{cb} = \frac{d_{23}d_{33}}{d_{32}^2+d_{33}^2}e^{i\varphi}\,\epsilon + \mathcal{O}(\epsilon^3)\,,\; V_{ts} = -\frac{d_{23}d_{33}}{d_{32}^2+d_{33}^2}e^{-i\varphi}\,\epsilon + \mathcal{O}(\epsilon^3)\,,\; V_{td} =\frac{u_{13}}{u_{33}}\epsilon^2 + \mathcal{O}(\epsilon^4)\,,\\[2mm]
    y_u^2 &\simeq \epsilon^{10}\,u_{11}^2\,,\; y_c^2 \simeq \epsilon^4\,u_{22}^2\,,\; y_t^2 \simeq u_{33}^2\,,\; y_d^2 \simeq \epsilon^{10}\,d_{11}^2\,,\;
    y_s^2 \simeq \epsilon^6\,\frac{d_{23}^2d_{32}^2}{d_{32}^2+d_{33}^2}\,, \; y_b^2 \simeq \epsilon^4\,(d_{32}^2 + d_{33}^2)\,.
\end{split}
\end{align}
so that,
\begin{align}
    \beta = {\rm arg}\!\left(-\frac{V_{cd}V_{cb}^*}{V_{td}V_{tb}^*}\right) = {\rm arg}\!\left(\frac{d_{13}d_{33}u_{33}}{(d_{32}^2+d_{33}^2)u_{13}}e^{-i\varphi}\right)\simeq -\varphi\,.
\end{align}
For $\varphi = -\pi/8$ we get $\beta \simeq \pi/8$, in agreement with the experimental measurement. At next-to-leading order, we find
\begin{align}
\begin{split}
    V_{tb} &= 1 - \frac{d_{23}^2d_{33}^2}{2 (d_{32}^2+d_{33}^2)^2}\epsilon^2 + \mathcal{O}(\epsilon^4)\,, \quad \; V_{cb} = -\frac{d_{23}d_{33}}{d_{32}^2+d_{33}^2}e^{i\varphi}\,\epsilon +  \frac{d_{23}^3d_{33}*(2d_{32}^2-d_{33}^2)}{2(d_{32}^2+d_{33}^2)^3}e^{i\varphi}\epsilon^3 + \mathcal{O}(\epsilon^5)\,,\\[2mm]
    V_{cd} &= -\frac{d_{13}}{d_{23}}\,\epsilon + \bigg[ \frac{d_{13}^3}{2d_{23}^3} - \frac{d_{13}\,d_{23}\,d_{33}^2}{2(d_{32}^2+d_{33}^2)^2} \bigg]\epsilon^3 + \mathcal{O}(\epsilon^5)\;, \quad \; V_{td} = \frac{u_{13}}{u_{33}}\epsilon^2 + \bigg[\frac{d_{13}d_{23}^2d_{33}}{(d_{32}^2+d_{33}^2)^2}e^{-i\varphi}-\frac{u_{13}d_{13}^2}{2u_{33}d_{23}^2} \bigg]\epsilon^4 + \mathcal{O}(\epsilon^6)\,,\\[2mm]
    \beta &= {\rm arg}\!\left(-\frac{V_{cd}V_{cb}^*}{V_{td}V_{tb}^*}\right) = {\rm arg}\!\left(\bigg[\frac{d_{13}\,d_{33}\,u_{33}}{u_{13}(d_{32}^2+d_{33}^2)} + \mathcal{O}(\epsilon^2)\bigg]\,e^{-i\varphi} - \frac{d_{13}^2d_{23}^2d_{33}^2u_{33}^2}{u_{13}^2(d_{32}^2+d_{33}^2)^3}\,e^{-i2\varphi}\epsilon^2 + \mathcal{O}(\epsilon^3)\right)\\
    &\simeq \frac{\pi}{8}-\frac{1}{2}\sqrt{2-\sqrt{2}} \frac{d_{13}d_{23}^2d_{33}u_{33}}{(d_{32}^2+d_{33}^2)^2u_{13}}\epsilon^2\simeq  \frac{\pi}{8}-\frac{1}{2}\sqrt{2-\sqrt{2}} \bigg|\frac{V_{us}}{V_{td}}\frac{y_s^2}{y_b^2}\bigg|.
\end{split}
\end{align}

\subsubsection*{Example III}
Finally, we analyze a pair of textures with $\varphi = 5\pi/8$. The Yukawa matrices are
\begin{align}
\begin{split}
    Y_u = \begin{pmatrix}
    u_{11}\epsilon^5 & 0 & 0\\
    0 & u_{22}\,\epsilon^2 & u_{23}\,\epsilon\\
    0 & 0 & u_{33}\\
    \end{pmatrix}\,,\quad
    Y_d = \begin{pmatrix}
    0 & e^{i\varphi}d_{12}\,\epsilon^4 & d_{13}\epsilon^4\\
    0& d_{22}\,\epsilon^3 & 0\\
    d_{31}\epsilon^3 & 0 & d_{33}\,\epsilon^2\\
    \end{pmatrix}\,.
\end{split}
\end{align}
$V_L^d$ is well-approximated by the product of a $1$-$3$ and a $1$-$2$ rotation, $V_L^d \simeq R_{12}\cdot R_{13}$, with $s^d_{13}\simeq -d_{13}/d_{33}\,\epsilon^2$ and $s^d_{12}=-e^{i\varphi}d_{12}/d_{22}\, \epsilon$\\
\begin{align}
    V^d_L \simeq \begin{pmatrix}
    1 & -s^d_{12} & s^d_{13}\\[2mm]
    -s^{d\, *}_{12} & 1 & -s^d_{13}s^{d\, *}_{12}\\[2mm]
    -s^{u\, *}_{13} & 0 & 1
    \end{pmatrix}\,.
\end{align}
$V_L^u$ is an exact $2$-$3$ rotation, with $s^u_{23} \simeq -u_{23}/u_{33}\epsilon$.
The CKM elements and quark Yukawas at leading order are
\begin{align}
\begin{split}
    V_{ud} &\simeq V_{cs} \simeq V_{tb} \simeq 1+\mathcal{O}(\epsilon^2)\,,\; V_{us} = \frac{d_{12}}{d_{22}}\,e^{-i\varphi}\epsilon + \mathcal{O}(\epsilon^3)\,,\; V_{ub} = \frac{d_{13}}{d_{33}}\epsilon^2 + \mathcal{O}(\epsilon^4)\,,\\
    V_{cd} &= -\frac{d_{12}}{d_{22}}\,e^{i\varphi}\epsilon + \mathcal{O}(\epsilon^3)\,,\; V_{cb} = -\frac{u_{23}}{u_{33}}\,\epsilon + \mathcal{O}(\epsilon^3)\,,\; V_{ts} = \frac{u_{23}}{u_{33}}\,\epsilon + \mathcal{O}(\epsilon^3)\,,\; V_{td} = -\frac{u_{23}d_{12}}{u_{33}d_{22}}e^{i\varphi}\epsilon^2 - \frac{d_{13}}{d_{33}}\epsilon^2 + \mathcal{O}(\epsilon^4)\,,\\
    y_u^2 &\simeq \epsilon^{10}\,u_{11}^2\,,\; y_c^2 \simeq \epsilon^4\,u_{22}^2\,,\; y_t^2 \simeq u_{33}^2\,,\; y_d^2 \simeq \epsilon^{10}\,\frac{d_{13}^2d_{31}^2}{d_{33}^2}\,,\;
    y_s^2 \simeq \epsilon^6\,d_{22}^2\,, \; y_b^2 \simeq \epsilon^4\,d_{33}^2\,.
\end{split}
\end{align}
Here $V_{ub}$ and $V_{cb}$ are real while $V_{us}$ and $V_{cd}$ carry the CP phase, so $\varphi$ lands in the ratio $V_{ud}V_{ub}^*/V_{cd}V_{cb}^*$ and hence in $\gamma$. At leading order,
\begin{align}
    \gamma = {\rm arg}\!\left(-\frac{V_{ud}V_{ub}^*}{V_{cd}V_{cb}^*}\right) = {\rm arg}\!\left(-\frac{d_{13}\,d_{22}u_{33}}{u_{23}\,d_{33}\,d_{12}}\,e^{-i\varphi}\right)\simeq \pi - \varphi\,.
\end{align}
For $\varphi=5\pi/8$, we get $\gamma\simeq 3\pi/8$, in agreement with experiments. At next-to-leading order we obtain
\begin{align}
\begin{split}
V_{ud} &= 1 \;-\; \frac{d_{12}^{2}}{2\,d_{22}^{2}}\,\epsilon^{2} \;+\; \mathcal{O}(\epsilon^{4})\,, \quad \quad \quad \quad V_{cd} = -\frac{d_{12}}{d_{22}}\,e^{i\varphi}\Bigl[\epsilon-\frac{u_{23}^{2}}{2\,u_{33}^{2}}\epsilon^{3}-\frac{d_{12}^{2}}{2\,d_{22}^{2}}\epsilon^{3}\,\Bigr]+\frac{u_{23}d_{13}}{u_{33}d_{33}}\epsilon^3 + \;\mathcal{O}(\epsilon^{5})\,,\\[2mm]
V_{ub} &= \frac{d_{13}}{d_{33}}\,\epsilon^{2}+\frac{d_{13}d_{31}^2}{d_{33}^3}\epsilon^4\;+\;\mathcal{O}(\epsilon^{6})\,, \quad \quad V_{cb} = -\frac{u_{23}}{u_{33}}\,\epsilon \;+\; \frac{u_{23}^3}{2\,u_{33}^{3}}\,\epsilon^{3}\, +\; \mathcal{O}(\epsilon^{5})\,.\\[2mm]
\gamma &= {\rm arg}\!\left(-\frac{V_{ud}V_{ub}^*}{V_{cd}V_{cb}^*}\right) = {\rm arg}\!\left(-\frac{d_{13}\,d_{22}u_{33}}{d_{12}\,d_{33}\,u_{23}}\,e^{-i\varphi} - \frac{d_{22}^2d_{13}^2}{d_{12}^2d_{33}^2}\,e^{-i2\varphi}\epsilon^2\right)\simeq \frac{3\pi}{8} - \frac{1}{2}\sqrt{2+\sqrt{2}}\,\frac{d_{22}d_{13}u_{23}}{d_{12}d_{33}u_{33}}\epsilon^2\\
& \simeq \, \frac{3\pi}{8} - \frac{1}{2}\sqrt{2+\sqrt{2}}\,\bigg|\frac{V_{cb}\, V_{ub}}{V_{us}}\bigg|.
\end{split}
\end{align}

\section{The $\pi/4$ peak}

\label{app:Pi4}

As already mentioned in Sec.~\ref{app:U(1)9}, while the $\{\pi/2,\pi/8, 3\pi/8\}$ values of the rephasing invariant phase in the 9-link textures have a simple explanation in terms of their connection to the angles of the unitarity triangle, the same is not true for the $\pi/4$ peak, which is related to a coincidence at the level of the quark masses. In this section, we focus on a particular example of a texture where $\varphi = \pi/4$ is needed to reproduce the data,
\begin{equation}
    \begin{gathered}
        \begin{tikzpicture}[scale=0.6,
        line cap=round, line join=round,
        vleft/.style={circle, fill=blue!65!black, draw=blue!65!black, inner sep=1.2pt},
        vright/.style={circle, draw=blue!65!black, fill=white, line width=0.8pt, inner sep=1.2pt},
        vstar/.style={star, star points=5, star point ratio=1.3, draw=black, fill=black, inner sep=1.2pt},
        link/.style={draw=gray!65!white, line width=0.9pt},
        linkr/.style={draw=red!50!black, line width=0.9pt},
        lab/.style={text=blue!25!black, font=\normalsize}
        ]
        \node[vleft]  (u1) at (0, 1) {};
        \node[vleft]  (u2) at (0, 0) {};
        \node[vleft]  (u3) at (0,-1) {};

        \node[vstar]  (q1) at (1.45, 1) {};
        \node[vstar]  (q2) at (1.45, 0) {};
        \node[vstar]  (q3) at (1.45,-1) {};

        \node[vright] (d1) at (2.90, 1) {};
        \node[vright] (d2) at (2.90, 0) {};
        \node[vright] (d3) at (2.90,-1) {};

        \node[lab] at (0,1.45) {$u^c$};
        \node[lab] at (1.45,1.45) {$q$};
        \node[lab] at (2.90,1.45) {$d^c$};

        \draw[link] (u1) -- (q1);
        \draw[link] (u2) -- (q2);
        \draw[link] (u3) -- (q2);
        \draw[link] (u3) -- (q3);

        \draw[link]  (q1) -- (d3);
        \draw[link]  (q2) -- (d1);
        \draw[linkr] (q2) -- (d2);
        \draw[link]  (q3) -- (d2);
        \draw[link]  (q3) -- (d3);

         \node[vleft]  (u1) at (0, 1) {};
        \node[vleft]  (u2) at (0, 0) {};
        \node[vleft]  (u3) at (0,-1) {};

        \node[vstar]  (q1) at (1.45, 1) {};
        \node[vstar]  (q2) at (1.45, 0) {};
        \node[vstar]  (q3) at (1.45,-1) {};

        \node[vright] (d1) at (2.90, 1) {};
        \node[vright] (d2) at (2.90, 0) {};
        \node[vright] (d3) at (2.90,-1) {};
    \end{tikzpicture}
    \end{gathered} \quad \Rightarrow  \quad  Y_u = \begin{pmatrix}
    u_{11}\,\epsilon^6 &  0 &0\\
    0 & u_{22}\,\epsilon^3 & u_{23}\,\epsilon^2\\
    0 & 0 & u_{33}\\
    \end{pmatrix}\,,\quad
    Y_d = \begin{pmatrix}
    0 & 0 & d_{13}\,\epsilon^5\\
    d_{21}\,\epsilon^5 & d_{22} \,\epsilon^4 e^{i \varphi} & 0\\
    0 & d_{32}\,\epsilon^2 & d_{33}\,\epsilon^2\\
    \end{pmatrix}\,,.
\end{equation}
Perturbatively diagonalizing this matrix we find that $Y_u$ can be diagonalized at leading order with a single rotation: $V^u_L \simeq R_{23}$ with $s_{23}^u = -u_{23}/u_{33}\, \epsilon^2$, while for $Y_d$: 
\begin{equation}
\begin{aligned}
    &V_L^d \simeq R_{12}\cdot R_{13}\cdot R_{23}  \simeq \begin{pmatrix}
    1 & s^d_{12} & s^d_{13}+s^d_{23}s^d_{12}\\
    -s^{d\,*}_{12} & 1 & s^d_{23}\\
    -s^{d\,*}_{13} & -s^{d\,*}_{23} & 1\\
    \end{pmatrix}\, , \\[1em]
&\text{with } \, \, s^d_{12} = \frac{d_{13} d_{32} e^{-i \varphi}}{d_{22} d_{33}}\epsilon, \quad  s^d_{13} = -\frac{d_{13} d_{33}}{d_{32}^2 + d_{33}^2} \epsilon^3, \quad s^d_{23} = -\frac{d_{22} d_{32} e^{i \varphi}}{d_{32}^2 + d_{33}^2} \epsilon^2\, .
\end{aligned}
\end{equation}
At leading order in $\epsilon$, CKM elements and quark masses are
\begin{align}
\begin{split}
    V_{ud} &\simeq V_{cs} \simeq V_{tb} \simeq 1 + \mathcal{O}(\epsilon^2)\,,\; V_{us} = -\epsilon\,\frac{d_{13}d_{32}}{d_{22}d_{33}}e^{-i\varphi} + \mathcal{O}(\epsilon^3)\,,\; V_{ub} = \epsilon^3\,\frac{d_{13}d_{33}}{d_{32}^2+d_{33}^2} + \mathcal{O}(\epsilon^4)\,,\; V_{cd} = \epsilon\,\frac{d_{13}d_{32}}{d_{22}d_{33}}e^{i\varphi} + \mathcal{O}(\epsilon^3)\,,\\
    V_{cb} &=  \epsilon^2\frac{d_{22}d_{32}u_{33}e^{i\varphi} - (d_{32}^2+d_{33}^2)u_{23}}{(d_{32}^2+d_{33}^2)u_{33}} + \mathcal{O}(\epsilon^4)\,,\; V_{td} = \epsilon^3\,\frac{d_{13}(d_{32}u_{23}e^{i\varphi}-d_{22}u_{33})}{d_{33}d_{22}u_{33}} + \mathcal{O}(\epsilon^4)\,,\\
    V_{ts} &= \epsilon^2\frac{(d_{32}^2+d_{33}^2)u_{23} - d_{22}d_{32}u_{33}e^{-i\varphi}}{(d_{32}^2+d_{33}^2)u_{33}} + \mathcal{O}(\epsilon^4)\,, + \mathcal{O}(\epsilon^3)\,,\;
    y_u^2 \simeq \epsilon^{12}u_{11}^2\,,\;  y_c^2 \simeq \epsilon^{6} u_{22}^2\,,\; y_t^2 \simeq u_{33}^2\,,\\
    y_d^2 &\simeq \epsilon^{12} \frac{d_{13}^2d_{21}^2d_{32}^2}{d_{22}^2d_{33}^2}\,,\; y_s^2 \simeq \epsilon^8 \frac{d_{22}^2d_{33}^2}{d_{32}^2+d_{33}^2}\,,\; y_b^2 \simeq \epsilon^{4} (d_{32}^2+d_{33}^2)\,.
\end{split}
\label{eq:LO_pi4}
\end{align}
Remembering the definition of $\alpha$, $\beta$ and $\gamma$, it is clear that none of them equal $\varphi$. Now to understand why $\varphi \sim -\pi/4$ is a solution to the fit, let us constrain its value from the features of the unitarity triangle which have to be reproduced by the fit. First of all, the fact that $\alpha \sim \pi/2$ implies that at leading order we should have: 
\begin{equation}
    |V_{cd} V_{cb}^* |^2 = |V_{ud} V_{ub}^*|^2 + |V_{td} V^*_{tb}|^2  \quad \Rightarrow \quad \cos{\varphi} = \frac{d_{22} u_{33}}{d_{32} u_{23}} \;.
\end{equation}
Then, the cosine law of the unitarity triangle implies
\begin{equation}
    \bigg|\frac{V_{td} V^*_{tb}}{V_{ud} V_{ub}^*}\bigg| = \tan{\gamma} \simeq \tan{\frac{3\pi}{8}} \quad \Leftrightarrow \quad | \tan{ \varphi} | \times \frac{d_{32}^2+d_{33}^2}{d_{33}^2} = \tan{\frac{3\pi}{8}}\,.
\end{equation}
Now using the leading expression in Eq.~\ref{eq:LO_pi4} we can solve for each $d_{ij}/u_{ij}$ in terms of the observables, and we find that: 
\begin{equation}
    \frac{d_{32}^2+d_{33}^2}{d_{33}^2}  \simeq 1 + \frac{y_s^2}{y_b^2} \left| \frac{V_{us}}{V_{ub}}\right|^2.
\end{equation}
Since $\tan{\frac{3\pi}{8}} = 1 + \sqrt{2}$, we conclude that in order for $\varphi = \pm \pi/4$ to be a solution we must have 
\begin{equation}
    \frac{y_s^2}{y_b^2} \left| \frac{V_{us}}{V_{ub}}\right|^2 \sim \sqrt{2}\,,
\end{equation}
which is correct at the $9\%$ level experimentally.

\section{Fit of 9-link textures}\label{app:Fit}

In this appendix, we describe the details of the different numerical analyses carried out in the main text. We start by explaining the details of the scan over all possible $9$-link textures, as well the identification of  textures within the same equivalence class. We then proceed to describe how the various constrained fits were performed, as well as the respective equivalence classes associated with each of them. 

\subsection{Scanning over $9$-link textures: 10 parameters scan}
\label{app:fit-free}
\begin{table}[t]
\centering
\setlength{\tabcolsep}{12pt}
\begin{tabular}{lcc}
\hline\hline
Observable & Central value & $\sigma$\\
\hline
$y_u$        & $7.04\times10^{-6}$ & $0.15\times10^{-6}$ \\
$y_c$        & $3.56\times10^{-3}$ & $0.06\times10^{-3}$ \\
$y_t$        & $0.967$             & $0.004$             \\
$y_d$        & $1.54\times10^{-5}$ & $0.02\times10^{-5}$ \\
$y_s$        & $3.06\times10^{-4}$ & $0.04\times10^{-4}$ \\
$y_b$        & $1.630\times10^{-2}$& $0.009\times10^{-2}$ \\
\hline
$|V_{us}|$   & $0.22517$  & $0.00068$ \\
$|V_{ub}|$   & $0.003763$ & $0.000088$\\
$|V_{cb}|$   & $0.04189$  & $0.00081$ \\
$|V_{cd}|$   & $0.22503$  & $0.00068$ \\
$|V_{td}|$   & $0.00863$  & $0.00019$ \\
$|V_{ts}|$   & $0.04117$  & $0.00079$ \\
\hline
$\alpha$     & $84.1^{\circ}$ & $3.7^{\circ}$ \\
$\beta$      & $22.6^{\circ}$ & $0.5^{\circ}$ \\
$\gamma$     & $66.4^{\circ}$ & $2.8^{\circ}$ \\
\hline
$y_u/y_d$                    & $0.473$ & $0.017$ \\
$y_s/\bar{y}_{ud}$           & $27.30$ & $0.08$  \\
\hline\hline
\end{tabular}
\caption{Central values and $1\sigma$ uncertainties at $M_Z$ (see Refs.~\cite{ParticleDataGroup:2026aaa,Antusch:2025fpm}) used to define
the 17-observable $\chi^2$ of Eq.~(1). Here $\bar{y}_{ud}\equiv(y_u+y_d)/2$. Note that the quoted $\alpha$ and $\gamma$ uncertainties are the larger of each observable's asymmetric $1\sigma$ errors, used symmetrically as a conservative simplification.}
\label{tab:targets}
\end{table}

We start by considering all possible 9-link textures--pairs of full-rank $(Y_u,Y_d)$ matrices with nine nonzero entries, with a single closed loop carrying the one rephasing invariant.  The nine entries are then split as $4{+}5$, $5{+}4$, $6{+}3$, or $3{+}6$ between $Y_u$ and $Y_d$. Since we are
looking for hierarchical matrices we also require entries $(3,3)$ (in both matrices) to be non-zero. As for the closed loop, it can have $4$ or $6$-links that are either fully contained in a single matrix, or that span across both matrices.
Given a closed loop, we can compute the rephasing invariant by assigning an orientation to the loop -- arrows for the links going around it -- and taking the product of the loop-entries; a given link appears conjugated if its arrow points towards a $q$ node (see Fig. \ref{fig:IntroFigure}, left). The argument of this monomial defines the phase, $\varphi$. Of course, given a closed loop there are two possible orientations, which simply differ by $\varphi \to -\varphi$. 
Finally, all nonzero entries of a texture are taken to be real and positive, except for one of the loop entries -- which carries the single physical CP phase, $\exp{i \varphi}$ \footnote{any phase can always be removed by a field redefinition of the quark fields, except for the physical CP phase which is confined to the rephasing-invariant}. Of course, the phase can be placed anywhere in the closed loop, but for definiteness in defining our matrices, the phase-carrying entry is chosen by a fixed rule: preferentially the down-type entry with smallest row and column index; if the loop contains no down-type entry, the smallest up-type entry instead. 

To perform the full scan, for every texture we let each entry (or its magnitude, for the case of the entry with the phase) vary between zero and roughly three orders of magnitude above its natural value, $Y_{ij}\sim\sqrt{m_i m_j}/v$, with $m_i$ the corresponding up- or down-type quark mass and $v$ the electroweak vev. As for the phase, since $\varphi$ is periodic, the domain $(-\pi,\pi]$ is partitioned into windows of width $\pi/8$, each optimized independently -- so that a texture, in general, has different viable fits, coming from the different phase windows. 

To fit each texture to the data, we minimize a Gaussian $\chi^2$ over 17 observables: the three up-type and three down-type Yukawa singular values, the six CKM magnitudes $|V_{ij}|$, the three unitarity-triangle angles $\alpha$, $\beta$, and $\gamma$, and the mass ratios $y_u/y_d$ and $y_s/(\frac{y_u+y_d}{2})$. The central values and uncertainties at $M_Z$ are given in Tab.~\ref{tab:targets}. We classify a texture as \emph{viable} if its minimum $\chi^2$ lies within the $3\,\sigma$ confidence region.

We find a total of 2398 viable fits, and computing the respective rephasing invariant phase in each fit we obtain the distribution shown on the left of Fig.~\ref{fig:HistogramDefSM}. As mentioned earlier, in defining the rephasing invariant there are two possible orientations which differ by complex-conjugation; so to produce the histogram in Fig.~\ref{fig:HistogramDefSM}, given a viable fit we extract the phase, $\varphi$, and record the value between $\pm \varphi$ which lies $[0,\pi]$ -- this is equivalent to picking the orientation of the rephasing invariant such that the phase lies $[0,\pi]$, and henceforth this choice is what we will call $\varphi$. 

One of the central results of this work, discussed extensively in the main text, is that $\varphi$ can only take discrete values that are integer multiples of $\pi/8$. An explanation of this phenomena has been discussed in previous appendices. This feature is not generic, but rather a consequence of the observed pattern of flavor parameters, and in particular of the measured CKM unitarity triangle angles. We can see this by repeating the scan for a ``deformed'' version of the Standard Model. Specifically, we artificially double the bottom-quark mass and modify the CKM matrix, while preserving unitarity, such that
$ (\alpha,\beta,\gamma) =(100^\circ,\,5^\circ,\,75^\circ)$. The results are shown in Fig.~\ref{fig:HistogramDefSM}, where a qualitatively different set of allowed values for $\varphi$ is obtained. In particular the $\pi/8, 3\pi/8$ and $\pi/2$ peaks do not appear anymore, consistently with our understanding of their origin in connection to the measured values of $\alpha$, $\beta$ and $\gamma$.\\
We emphasize that while our scan tries to be as complete as possible it certainly does not find \textit{all} possible viable texture fits. For a given texture, our scan looks for the best possible $\chi^2$ solution, and therefore we might miss some viable local minima. For example, given a viable fit from our scan, corresponding to some texture, $T_a$, we can of course perform arbitrary row/column permutations -- inside the $S_3^Q \times S_3^{u^c} \times S_3^{d^c}$ -- to find another viable fit for a different texture, $T_b$, but which has a worse $\chi^2$ compared to the original one found in the scan for this texture; in such cases our scan does not keep track of both solutions. The invariant data is therefore given by the corresponding equivalence classes of viable fits where we mod-out by row/column permutations. In practice, we identify two texture fits, $T_a$ and $T_b$, if there is a permutation that maps $T_a$ to $T_b$, and such that the absolute value of all permuted entries of $T_a$ matches those of $T_b$ up to $5 \%$ and $|\varphi|$ agrees up to $0.1^\circ$. This leaves us with a total of $156$ classes.

Finally, from the analysis carried out in the previous appendices, it is clear that the mechanism that ensures that $\varphi= \pi/8$ is a possible fit is the same as that for $\varphi= 7\pi/8$ -- up to a minus sign, both yield the leading approximation for $\beta$; and similar for the $3\pi/8$ and $5\pi/8$ peaks. For this reason, in Fig.\ref{fig:theta_histogram} of the main text, we decided to present the ``folded version'' of the histogram above, where we present the results after modding out by the $S_3^3$ and we further identify $\varphi \to \varphi + \pi$. 

\subsection{Fixed Phase Fit}
\label{app:fit-fixed}

We now turn to the study of 9-link textures with the rephasing-invariant phase fixed to an integer multiple of $\pi/8$, and investigate the resulting predictions for the unitarity-triangle angles. The parametrization, texture generation, loop selection, and $\chi^2$ definition are identical to those described in App.~\ref{app:fit-free}; the only difference is that $\varphi$ is fixed rather than fitted, leaving 9 free magnitude parameters. In practice, for each texture we run the constrained fit, where the phase is pinned at $\varphi= \{\pm \pi/8,\ \pm 3\pi/8,\ \pm \pi/2,\ \pm 5\pi/8,\ \pm 7\pi/8 \}$. The case $\varphi=\pm\pi/4$ was discussed separately in App.~\ref{app:Pi4} and is not considered here, as it is unrelated to the observed values of the unitarity-triangle angles. For the purposes of our discussion, as mentioned above, we will not distinguish between $\pm\varphi$ or $\pi\pm\varphi$ --  and so refer to the different fit cases as $\varphi=\pi/8,\ 3\pi/8,\ \pi/2$, with the understanding that the corresponding rephasing-invariant phase may equally well be negative or shifted by $\pi$. 

From this fixed-phase scan we find a total of 1412 viable fits. After dividing these into their respective equivalence classes,  which identify textures related by permutations of the quark fields,  we find:
\begin{figure}[t]
    \centering
    \includegraphics[width=\linewidth]{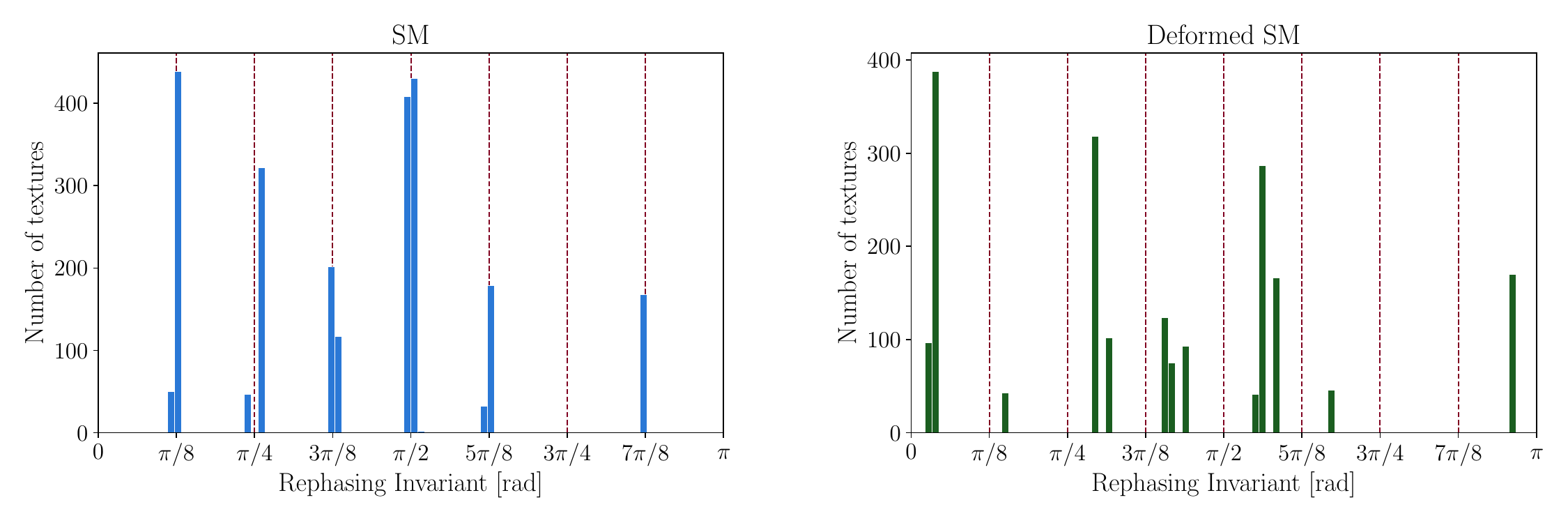}
    \caption{\textbf{(Left)} Histogram of the rephasing-invariant phase $\varphi$ (up to complex conjugation), showing the raw number of fits for the Standard Model and \textbf{(right)} for a deformed version in which the bottom-quark mass is doubled and the unitarity-triangle angles are fixed to $\alpha=100^\circ$, $\beta=5^\circ$, and $\gamma=75^\circ$.}
    \label{fig:HistogramDefSM}
\end{figure}
\begin{table}[h]
\centering
\begin{tabular}{c|ccc}
$\varphi$ & \hspace{0.2cm}$\pi/2$\hspace{0.4cm} & $\pi/8$ \hspace{0.6cm}&  \hspace{0.15cm} $3\pi/8$ \hspace{0.2cm}\\
\hline
No. Equiv. Classes \hspace{0.2cm}& 29 \hspace{0.4cm}& 35 \hspace{0.6cm}& \hspace{0.15cm} 35\hspace{0.2cm}\\
\end{tabular}
\end{table}

The next step is then to find the most \textit{hierarchical} element in each permutation class. To do this, we scan through the whole permutation orbit and ask for an element where both the left- and right-diagonalizing matrices are ``hierarchical", which we define to mean that their diagonal entries all have magnitude $\geq 0.95$. As it turns out 23/29 ($\pi/2$), 25/35 ($\pi/8$), and 26/35 ($3\pi/8$) classes have precisely one element satisfying the criterion above. As for the remaining classes, there is always one element that obeys the criterion for the left-diagonalizing matrix, but still has somewhat large mixing angles on the right. In Figs.~\ref{fig:29textures_pi2}, \ref{fig:35textures_pi8} and \ref{fig:35textures_3pi8}, we present the  link-diagrams for all these classes, where for a given class we present the texture corresponding to the most hierarchical element; we mark with $(\dagger)$ the textures which do not satisfy the hierarchical criterion on the right.

As pointed out in the main text, these equivalence classes are no longer related by the full $U(3)^3$ flavor symmetry: once the rephasing-invariant phase of the theory is fixed to $\pi/2$, $\pi/8$, or $3\pi/8$, a generic $U(3)^3$ transformation no longer preserves it. 
Nonetheless, beyond the permutation group, $S_3^3$, there is still a residual set of rotations that preserve the rephasing-invariant phase which we should mod out by to correctly identify theories which share an identical low-energy phenomenology. In practice this can be done by either identifying the rotations that map one fit into another; or by looking into textures where the observables of the best possible fit match up to very high-order -- that converge to the same minimum to about 10 significant figures in $\chi^2$. 

After modding out by this residual rotation group we are left with  9, 10, and 13 inequivalent classes for $\varphi=\pi/2$, $\pi/8$, and $3\pi/8$, respectively. The  textures belonging to each of these $U(3)$-fixed phase classes are reported in the left tables in Figs.~\ref{fig:29textures_pi2}- \ref{fig:35textures_3pi8}. 

Picking again the most hierarchical element of each class, and applying the perturbative diagonalization described in App. \ref{app:Examples}, we derive the leading order result for $R_{\alpha}$, $R_{\beta}$ or $R_{\gamma}$ as shown in Tab.~\ref{tab:LeadingOrderAngles}. To begin we note that for $\varphi = 3\pi/8$, there are 3 classes where $\gamma \sim 3\pi/8$ is an accident rather than a structural consequence of the Yukawa texture -- these are labeled by a, b and c. Therefore, we neglect these theories in our discussion, leaving us with $10$ classes for $\varphi = 3\pi/8$. In Fig.4 of the main text we show the angle predictions  for all the $(9,10,10)$ classes.

One final point worth highlighting is that in most cases the lack of a hierarchical structure on the right leads to a $R_{\alpha,\beta,\gamma}$ which is \textit{not} given by a single monomial of the Yukawa entries. For example, in the $\varphi=\pi/2$, the $U(3)$ groups $2$ and $9$ are made out of textures from Fig. \ref{fig:29textures_pi2}, which are marked with $(\dagger)$; the same holds for groups 8 and 10 in the $\varphi=\pi/8$ case; and in the $\varphi=3\pi/8$ all permutation textures which are \textit{not} hierarchical on the right precisely fall into the $a,b,c$ groups once we mod-out by the remaining rotations! So we see that all groups in which the leading contribution to the rephasing invariant is given by a single monomial are precisely those that contain at least one fully hierarchical texture in it's equivalence class.

\begin{table}[t]
\centering
\renewcommand{\arraystretch}{2.3}
\hspace{-1.1cm}
\begin{minipage}[t]{0.28\textwidth}
\centering
\begin{tabular}{|c|c|}
\hline
Tex. & $R_\alpha = -\dfrac{V_{td}\, V_{tb}^*}{V_{ud}\, V_{ub}^*}$ \\
\hline
$1$ & $\dfrac{u_{23}\, u_{32}\, d_{12}}{u_{12}\, u_{33}\, d_{22}}\; e^{i\pi/2}$ \\[0.2cm]
\hline
$2$ & $\dfrac{u_{23}\,(d_{32}^{2} + d_{33}^{2})}{u_{33}\, d_{22}\, d_{32}}\; e^{i\pi/2}$ \\[0.2cm]
\hline
$3$ & $\dfrac{d_{12}\, u_{22}}{d_{22}\, u_{12}}\; e^{i\pi/2}$ \\[0.2cm]
\hline
$4$ & $\dfrac{d_{12}\, u_{22}}{d_{22}\, u_{12}}\; e^{i\pi/2}$ \\[0.2cm]
\hline
$5$ & $\dfrac{u_{13}\, u_{32}}{u_{33}\, u_{12}}\; e^{i\pi/2}$ \\[0.2cm]
\hline
$6$ & $\dfrac{u_{13}\, u_{32}}{u_{33}\, u_{12}}\; e^{i\pi/2}$ \\[0.2cm]
\hline
$7$ & $\dfrac{u_{23}\, d_{33}}{u_{33}\, d_{23}}\; e^{i\pi/2}$ \\[0.2cm]
\hline
$8$ & $\dfrac{u_{23}\, d_{33}}{u_{33}\, d_{23}}\; e^{i\pi/2}$ \\[0.2cm]
\hline
$9$ & $\dfrac{u_{23}\,(d_{32}^{2} + d_{33}^{2})}{u_{33}\, d_{22}\, d_{32}}\; e^{i\pi/2}$ \\[0.2cm]
\hline
\end{tabular}
\end{minipage}
\hspace{-0.4cm}
\begin{minipage}[t]{0.28\textwidth}
\centering
\begin{tabular}{|c|c|}
\hline
Tex. & $R_\beta = -\dfrac{V_{cd}\, V_{cb}^*}{V_{td}\, V_{tb}^*}$ \\[0.2cm]
\hline
$1$ & $\dfrac{u_{12}\, d_{23}}{u_{22}\, d_{13}}\; e^{i\pi/8}$ \\[0.2cm]
\hline
$2$ & $\dfrac{u_{12}\, d_{23}\, u_{33}}{u_{22}\, d_{33}\, u_{13}}\; e^{i\pi/8}$ \\[0.2cm]
\hline
$3$ & $\dfrac{u_{12}\, u_{23}}{u_{22}\, u_{13}}\; e^{i\pi/8}$ \\[0.2cm]
\hline
$4$ & $\dfrac{u_{12}\, u_{23}\, d_{33}}{u_{22}\, u_{33}\, d_{13}}\; e^{i\pi/8}$ \\[0.2cm]
\hline
$5$ & $\dfrac{u_{12}\, u_{23}}{u_{22}\, u_{13}}\; e^{i\pi/8}$ \\[0.2cm]
\hline
$6$ & $\dfrac{u_{32}\, d_{23}}{u_{22}\, d_{33}}\; e^{i\pi/8}$ \\[0.2cm]
\hline
$7$ & $\dfrac{u_{12}\, d_{33}}{u_{32}\, d_{13}}\; e^{i\pi/8}$ \\[0.2cm]
\hline
$8$ & $\dfrac{d_{13}\, d_{33}\, u_{33}}{(d_{32}^2+d_{33}^2)\, u_{13}}\; e^{i\pi/8}$ \\[0.2cm]
\hline
$9$ & $\dfrac{u_{12}\, d_{23}}{u_{22}\, d_{13}}\; e^{i\pi/8}$ \\[0.2cm]
\hline
$10$ & $\dfrac{d_{13}\, d_{33}\, u_{33}}{(d_{32}^2+d_{33}^2)\, u_{13}}\; e^{i\pi/8}$ \\[0.2cm]
\hline
\end{tabular}
\end{minipage}
\hspace{0.13cm}
\begin{minipage}[t]{0.33\textwidth}
\centering
\begin{tabular}{|c|c|}
\hline
Tex. & $R_\gamma = -\dfrac{V_{ud}\, V_{ub}^*}{V_{cd}\, V_{cb}^*}$ \\[0.2cm]
\hline
 
$1$ & $\dfrac{d_{13}\, d_{22}}{d_{23}\,d_{12}}\, e^{i\,3\pi/8}$ \\[0.2cm]
\hline
 
$2$ & $\dfrac{u_{13}\, d_{22}\, d_{33}}{u_{33}\, d_{12}\, d_{23}}\; e^{i\,3\pi/8}$ \\[0.2cm]
\hline
 
$3$ & $\dfrac{d_{13}\, d_{22}}{d_{12}\, d_{23}}\; e^{i\,3\pi/8}$ \\[0.2cm]
\hline
 
$4$ & $\dfrac{d_{13}\, d_{22}\, u_{33}}{d_{33}\, d_{12}\, u_{23}}\; e^{i\,3\pi/8}$ \\[0.2cm]
\hline
 
$5$ & $\dfrac{u_{13}\, d_{22}}{d_{12}\, u_{23}}\; e^{i\,3\pi/8}$ \\[0.2cm]
\hline
 
$6$ & $\dfrac{u_{13}\, d_{22}}{d_{12}\, u_{23}}\; e^{i\,3\pi/8}$ \\[0.2cm]
\hline
 
$7$ & $\dfrac{u_{22}\, u_{33}}{u_{23}\, u_{32}}\; e^{i\,3\pi/8}$ \\[0.2cm]
\hline
 
$8$ & $\dfrac{u_{22}\, u_{33}}{u_{23}\, u_{32}}\; e^{i\,3\pi/8}$ \\[0.2cm]
\hline
 
$9$ & $\dfrac{d_{13}\, u_{33}}{u_{13}\, d_{33}}\; e^{i\,3\pi/8}$ \\[0.2cm]
\hline
 
$10$ & $\dfrac{d_{13}\, u_{33}}{u_{13}\, d_{33}}\; e^{i\,3\pi/8}$ \\[0.2cm]
\hline

a & $\dfrac{d_{32}\,\bigl(d_{13}\, d_{33} - d_{12}\, d_{32}\, e^{-i\,3\pi/8}\bigr)}{d_{33}\,\bigl(d_{13}\, d_{32} + d_{12}\, d_{33}\, e^{-i\,3\pi/8}\bigr)}$ \\[0.2cm]
\hline

b & $\dfrac{d_{32}\,\bigl( u_{12}\, d_{23}\, d_{33} - u_{22}\, d_{12}\, d_{32}\, e^{-i\,3\pi/8}\bigr)}{d_{33}\,\bigl(u_{12}\, d_{23}\, d_{32} + u_{22}\, d_{12}\, d_{33}\, e^{-i\,3\pi/8}\bigr)}$ \\[0.2cm]
\hline
 
c & $\dfrac{d_{32}\,\bigl(d_{13}\, d_{33} + d_{12}\, d_{32}\, e^{i\,5\pi/8}\bigr)}{d_{33}\,\bigl(d_{13}\, d_{32} - d_{12}\, d_{33}\, e^{i\,5\pi/8}\bigr)}$ \\[0.2cm]
\hline

\end{tabular}
\end{minipage}

\caption{Leading-order expressions for the unitarity-triangle ratios $R_\alpha$, $R_\beta$ and $R_\gamma$ for the $\pi/2$, $\pi/8$ and $3\pi/8$ $U(3)^3$ equivalence classes, whose textures and predictions are shown in Fig.~4 of the main text. Note that $\gamma \sim 3\pi/8$ is accidental rather than structural in the last three $3\pi/8$ classes and these have therefore been ignored in our analysis.
\label{tab:LeadingOrderAngles}}
\vspace{-0.2cm}
\end{table}

\subsection{Yukawa Triangle Fit}
We now investigate 9-link textures that predict the full CKM unitarity triangle, by imposing the Yukawa triangle to have angles$=(\pi/2,\pi/8,3\pi/8)$. As discussed in the main text, this can be achieved by fixing not only the phase of the rephasing-invariant monomial, but also its magnitude. For example, in textures for which $R_{\alpha}=V_{td}V_{tb}^*/V_{ud}V_{ub}^*$ is given at leading order by a single rephasing-invariant monomial, the angle is determined through $\alpha \simeq \arg(-R_{\alpha})$, while $|R_{\alpha}|$ fixes the ratio of the two edges of the unitarity triangle. 
Hence, specifying both the phase and the magnitude of $R_{\alpha}$ completely determines the shape of the triangle, up to overall scale. Similar considerations apply to $R_{\beta}$ and $R_{\gamma}$. 

Inspecting Tab.~\ref{tab:LeadingOrderAngles}, we find that, for the representatives of several equivalence classes, $R_{\alpha}$, $R_{\beta}$, or $R_{\gamma}$ is given at leading order by a single rephasing-invariant monomial. Consequently, fixing its magnitude amounts to fixing the relative magnitudes of a subset of Yukawa entries. Precisely those that belong to the rephasing-invariant loop defined at the level of the Yukawa matrices. As explained in the main text, this is what ultimately motivates us to define the ``Yukawa triangle'' whose sides are proportional to the different ratios entering in this monomial.

On the other hand, we also saw that there are classes in which $R_{\alpha}$, $R_{\beta}$, or $R_{\gamma}$ is not given by a single monomial at leading order -- associated to classes containing exclusively elements which are not hierarchical on the right.  For these, it is no longer the case that the unitarity triangle is, at leading order, the Yukawa triangle, and thus it does not make sense to fix the full rephasing-invariant. Note that the results in table \ref{tab:LeadingOrderAngles} were derived for a single representative of each equivalence class -- yet these classes were defined under $U(3)^3$ transformations that preserve the phase of the rephasing invariant, but not, in general, its magnitude -- so it could be that for a different element of the same class one would indeed find a different leading order expression, given solely by the rephasing-invariant monomial. While this is logically possible, it seems unlikely because we always choose the most hierarchical  element of each class precisely to avoid competition between different entries. A more likely explanation is that the leading order results were derived from the numerical fits which found the minima for the $\chi^2$ when the phase was fixed; but it could be that once we demand that the full triangle is fixed the minima changes location and therefore so does the leading contributions to $R_{\alpha,\beta,\gamma}$. 

We therefore proceed as follows. First, we return to the $(29,35,35)$ permutation equivalence classes. For each representative texture, we identify the relevant rephasing invariant and repeat the fit to the flavor observables with both its phase and its magnitude held fixed. In practice, fixing the magnitude of the rephasing invariant amounts to expressing one Yukawa entry in terms of products and ratios of other Yukawa entries. We find that $(23,27,33)$ textures provide a fit with a $\chi^2$ within the $3\sigma$ confidence region for $\varphi=\pi/2$, $\pi/8$, and $3\pi/8$, respectively. We can now mod out by the action of remaining $U(3)^3$ -- which preserves not only the phase but the full-rephasing-invariant -- thereby identifying theories with identical low-energy phenomenology.  We find $(17,19,19)$ classes for $\varphi=\pi/2$, $\pi/8$, and $3\pi/8$, respectively, whose members are shown in the bottom-right tables of Figs.~\ref{fig:29textures_pi2}, \ref{fig:35textures_pi8}, and \ref{fig:35textures_3pi8}.

Looking at ~\ref{fig:29textures_pi2}, we indeed find that the six $\varphi=\pi/2$ classes for which the fit failed (absent in the Phase and Triangle Fit table) are exactly those that are not hierarchical on the right. In the $\varphi=\pi/8$ textures the result is more mixed: while all classes that do not find a viable fit also fail to be  hierarchical on the right -- these are classes 2, 3, 4, 6, 13, 16, 21 and 24 --  some of the original non-hierarchical textures -- classes 12 and 23 -- converge to a new minima which is now hierarchical, and the fit succeeds!  In the $\varphi=3\pi/8$ textures, we see even more vividly point (2): all textures from groups $a$, $b$ and $c$ -- that failed the hierarchical criterion -- find a new minima in the constrained fit yielding fully hierarchical matrices. Only classes 9 and 13 fail to find good enough fits under this new constraint. 

The predictions in the $\alpha$--$\beta$ plane for the most hierarchical representative of each equivalence class are shown in Fig.~\ref{fig:ZoomIns} of the main text. Each point is labeled by the corresponding texture number from Figs.~\ref{fig:29textures_pi2}, \ref{fig:35textures_pi8}, and \ref{fig:35textures_3pi8}. We emphasize that the astonishing precision of these predictions, all accurate to better than $0.3^\circ$ for each unitarity-triangle angle, compared with those obtained by fixing only the leading-order contribution to a single angle, stems from the fact that we now determine the entire unitarity triangle at leading order, rather than just one of its angles.

\begin{figure}
    \centering
     \input{Figures/29Textures_pi2}
    \caption{Equivalence classes, under row/column permutations, for textures with $\varphi = \pi/2$. The most hierarchical representative for each class is shown. The equivalence classes under $U(3)^3$ rotations for fixed rephasing-invariant phase as well as fixed full Yukawa triangle are shown at the bottom. The link where the phase is placed (marked in red) is always such that Arg(Det$(Y_uY_d)$)=0; except when this is not possible, such cases are marked with $(*)$. The $(\dagger)$ denotes classes for which the right-diagonalizing matrix does not obey the hierarchical criterion.}
    \label{fig:29textures_pi2}
\end{figure}
\begin{figure}
    \centering
     \input{Figures/35Textures_pi8}
    \caption{Equivalence classes, under row/column permutations, for textures with $\varphi = \pi/8$. The most hierarchical representative for each class is shown. The equivalence classes under $U(3)^3$ rotations for fixed rephasing-invariant phase as well as fixed full Yukawa triangle are shown at the bottom. The link where the phase is placed (marked in red) is always such that Arg(Det$(Y_uY_d)$)=0; except when this is not possible, such cases are marked with $(*)$. The $(\dagger)$ denotes classes for which the right-diagonalizing matrix does not obey the hierarchical criterion.}
    \label{fig:35textures_pi8}
\end{figure}

\begin{figure}
    \centering
     \input{Figures/35Textures_3pi8}
    \caption{Equivalence classes, under row/column permutations, for textures with $\varphi = 3\pi/8$. The most hierarchical representative for each class is shown. The equivalence classes under $U(3)^3$ rotations for fixed rephasing-invariant phase as well as fixed full Yukawa triangle are shown at the bottom. The link where the phase is placed (marked in red) is always such that Arg(Det$(Y_uY_d)$)=0; except when this is not possible, such cases are marked with $(*)$. The $(\dagger)$ denotes classes for which the right-diagonalizing matrix does not obey the hierarchical criterion.}
    \label{fig:35textures_3pi8}
\end{figure}

\end{document}